\documentclass[longauth]{aa}  
\usepackage{ulem}
\usepackage{graphicx}
\usepackage{txfonts}
\usepackage{xcolor}
\usepackage{amsmath, amssymb} 
\usepackage[colorlinks=true,linkcolor=red,anchorcolor=blue,citecolor=blue,filecolor=black,menucolor=black,runcolor=black,urlcolor=blue]{hyperref}

\newcommand{\oddpm}[2]{\raisebox{0.5ex}{\tiny$\substack{+#1 \\ -#2}$}}

\newcommand{\Mstar}{$M_\mathrm{*}$\xspace}
\newcommand{\Msyr}{\,$M_\odot$\,yr$^{-1}$\xspace}

\newcommand{\arcs}{$\arcsec$\xspace}
\newcommand{\micron}{\,$\mu$m\xspace}
\newcommand{\Myr}{\,Myr\xspace}

\newcommand{\Msun}{\,$M_\mathrm{\odot}$\xspace}
\newcommand{\Mdyn}{\,$M_\mathrm{dyn}$\xspace}

\newcommand{\Lsun}{\,$L_\mathrm{\odot}$\xspace}
\newcommand{\Zsun}{\,$Z_\mathrm{\odot}$\xspace}

\newcommand{\kms}{\,km\,s$^{-1}$\xspace}

\newcommand{\ergs}{\,erg\,s$^{-1}$\xspace}
\newcommand{\cubiccm}{\,cm$^{-3}$\xspace}
\newcommand{\K}{\,K\xspace}
\newcommand{\ergscm}{\,erg\,s$^{-1}$\,cm$^{-2}$\xspace}

\newcommand{\EXP}[1]{\,$\times$\,10$^{#1}$}
\newcommand{\range}{\,$-$\,}
\newcommand{\eg}{e.g.\,}
\newcommand{\ie}{i.e.\,}
\newcommand{\eqq}{\,=\,}
\newcommand{\pmm}{\,$\pm$\,}
\newcommand{\simi}{$\sim$\,}
\newcommand{\z}{$z$\xspace}
\newcommand{\Te}{$T_\mathrm{e}$\xspace}
\newcommand{\Ne}{$n_\mathrm{e}$\xspace}
\newcommand{\nuclear}{`nuclear'\xspace}
\newcommand{\total}{`total'\xspace}
\newcommand{\Av}{$A_\mathbf{V}$\xspace}
\newcommand{\RXC}{RXCJ2248-ID\xspace}
\newcommand{\RXCthree}{RXCJ2248-ID3\xspace}

\newcommand{\HIn}{\ion{H}{I}\xspace}

\newcommand{\HeIn}{\ion{He}{I}\xspace}

\newcommand{\Ha}{H$\alpha$\xspace}
\newcommand{\Hb}{H$\beta$\xspace}
\newcommand{\Hg}{H$\gamma$\xspace}
\newcommand{\Hd}{H$\delta$\xspace}

\newcommand{\ArIV}[1]{[\ion{Ar}{IV}]\,$\lambda$#1}
\newcommand{\ArIVn}{[\ion{Ar}{IV}]\xspace}
\newcommand{\OIII}[1]{[\ion{O}{III}]\,$\lambda$#1}
\newcommand{\OIIIn}{[\ion{O}{III}]\xspace}
\newcommand{\OIIIalma}{[\ion{O}{III}]\,88$\mu$m\xspace}

\newcommand{\HeI}[1]{\ion{He}{I}\,$\lambda$#1}
\newcommand{\NeIV}[1]{[\ion{Ne}{IV}]\,$\lambda$#1}
\newcommand{\NII}[1]{[\ion{N}{II}]\,$\lambda$#1}

\begin{document}

    \title{RIOJA: The dusty outflows and density-complex interstellar medium in the N-enhanced lensed galaxy RXCJ2248-ID at \z{}\eqq6.1}

   \author{A. Crespo G\'omez\inst{\ref{inst:CAB},\ref{inst:STScI}}
   \and Y. Tamura\inst{\ref{inst:Nagoya}}
   \and L. Colina\inst{\ref{inst:CAB}}
   \and J. \'Alvarez-M\'arquez\inst{\ref{inst:CAB}}
   \and T. Hashimoto \inst{\ref{inst:Tsukuba}}
   \and R. Marques-Chaves\inst{\ref{inst:Geneva}}
   \and Y. Nakazato \inst{\ref{inst:DoPTokio},\ref{inst:Flatiron}}
   \and C. Blanco-Prieto\inst{\ref{inst:CAB}} 
   \and K. Sunaga\inst{\ref{inst:Nagoya}} 
   \and L. Costantin \inst{\ref{inst:CAB}}
   \and A. K. Inoue\inst{\ref{inst:SASE},\ref{inst:Waseda}}
   \and A. Hamada \inst{\ref{inst:Tsukuba}}
   \and S. Arribas \inst{\ref{inst:CAB}}
   \and D. Ceverino \inst{\ref{inst:UAM},\ref{inst:CIAFF}}
   \and M. Hagimoto \inst{\ref{inst:Nagoya}} 
   \and K. Mawatari \inst{\ref{inst:SASE},\ref{inst:Waseda}}
   \and W. Osone  \inst{\ref{inst:Tsukuba}} 
   \and Y. Sugahara \inst{\ref{inst:SASE},\ref{inst:Waseda}}
   \and N. Yoshida \inst{\ref{inst:DoPTokio},\ref{inst:UniTokyo_Kavli},\ref{inst:RCEU}}
   %\and Y. Ren \inst{\ref{inst:Shinjuku}} 
   \and Y. Harikane \inst{\ref{inst:UniTokyo}}
   \and M. M. Lee \inst{\ref{inst:DAWN}} 
   \and A. Taniguchi \inst{\ref{inst:Kitami}} 
   \and H. Umehata \inst{\ref{inst:Nagoya}} 
   }

    \institute{Centro de Astrobiolog\'{\i}a (CAB), CSIC-INTA, Ctra. de Ajalvir km 4, Torrej\'on de Ardoz, E-28850, Madrid, Spain \label{inst:CAB}
    \and Space Telescope Science Institute (STScI), 3700 San martin Drive, Baltimore, MD 21218, USA\\    \email{acrespo@stsci.edu} \label{inst:STScI}
    \and Department of Physics, Graduate School of Science, Nagoya University, Furo, Chikusa, Nagoya, Aichi 464-8602, Japan \label{inst:Nagoya}
    \and Division of Physics, Faculty of Pure and Applied Sciences, University of Tsukuba, Tsukuba, Ibaraki 305-8571, Japan \label{inst:Tsukuba}
    \and Geneva Observatory, Department of Astronomy, University of Geneva, Chemin Pegasi 51, CH-1290 Versoix, Switzerland\label{inst:Geneva}
    \and Department of Physics, The University of Tokyo, 7-3-1 Hongo, Bunkyo, Tokyo 113-0033, Japan \label{inst:DoPTokio}
    \and Center for Computational Astrophysics, Flatiron Institute, 162 5th Avenue, New York, NY 10010 \label{inst:Flatiron}
    \and Department of Physics, School of Advanced Science and Engineering, Faculty of Science and Engineering, Waseda University, 3-4-1 Okubo, Shinjuku, Tokyo 169-8555, Japan \label{inst:SASE}
    \and Waseda Research Institute for Science and Engineering, Faculty of Science and Engineering, Waseda University, 3-4-1 Okubo, Shinjuku, Tokyo 169-8555, Japan \label{inst:Waseda}
    \and Departamento de Fisica Teorica, Modulo 8, Facultad de Ciencias, Universidad Autonoma de Madrid, 28049 Madrid, Spain \label{inst:UAM}
    \and  CIAFF, Facultad de Ciencias, Universidad Autonoma de Madrid, 28049 Madrid, Spain \label{inst:CIAFF}
    \and Kavli Institute for the Physics and Mathematics of the Universe (WPI), UT Institute for Advanced Study, The University of Tokyo, Kashiwa, Chiba 277-8583, Japan \label{inst:UniTokyo_Kavli}
    \and Research Center for the Early Universe, School of Science, The University of Tokyo, 7-3-1 Hongo, Bunkyo, Tokyo 113-0033, Japan \label{inst:RCEU}
    \and Institute for Cosmic Ray Research, The University of Tokyo, 5-1-5 Kashiwanoha, Kashiwa, Chiba 277-8582, Japan\label{inst:UniTokyo}
    \and Cosmic Dawn Center (DAWN), Niels Bohr Institute, University of Copenhagen, Jagtvej 128, DK-2200 Kobenhavn N, Denmark \label{inst:DAWN}
    \and Kitami Institute of Technology, 165 Koen-cho, Kitami, Hokkaido 090-8507, Japan \label{inst:Kitami}
    }   

   \date{Received ; accepted}

\abstract{We present an analysis on the kinematics and physical properties of the ionised gas in the lensed galaxy \RXC at \z{}\eqq6.105 based on high-resolution JWST NIRSpec/IFU data. This source is a nitrogen-rich high-\z galaxy showing a compact morphological structure and high star formation and stellar surface densities. The magnification factor of the observed image of the source (ID3, $\mu$\eqq5.3) along with the rich dataset available, including NIRSpec and ALMA spectroscopy, makes this target an ideal benchmark to study the physical properties of nitrogen-rich galaxies. In this work, we use the large number of optical lines detected at a high signal-to-noise (S/N) in our NIRSpec high-resolution (R\eqq2700) integral field spectra to analyse potential differences in kinematics, dust distribution, and electron temperature and density. In addition, we combine our data with available ALMA observations to derive the optical-to-far-IR \OIII{5008}/\OIIIalma ratio in order to estimate the electron density of the O$^{++}$ regions. Our analysis revealed 
a high electron temperature (\Te{}\,\simi30000\K) in the ionised gas independent of the ionisation level. We measured a wide range in the electron densities of the ionised interstelllar medium (ISM) from our \OIII{5008}/\OIIIalma and \ArIV{4713}/\ArIV{4742} ratios (log(\Ne{}[\cubiccm{}])\,\simi2.7\range3.8) and previous values (\ie 4.8\,$<$\,log(\Ne{}[\cubiccm{}])\,$<$\,5.5) based on high-ionisation rest-UV emission lines. The ionised gas appears to be clumpy with a low filling factor ranging from about 10\% to 0.2\% for the low- and high-density clouds. In addition, we observed a very complex ISM kinematic structure traced by the \OIII{5008} and \Ha lines. Concretely, we derived the presence of two distinct broad and very broad components (FWHM\,\simi210\range250 and \simi1000\range1500\kms) in addition to the systemic component (FWHM\,\simi60\range70\kms). These broad components are heavily extinct (\Av{}\,\simi1.5 and 2.5, respectively), based on their Balmer decrements, while the gas associated with the narrow component is consistent with no extinction. The maximal velocities of these outflows (\simi115\range500\kms) are such that a fraction of the total outflowing gas (0.16\range2.1\EXP{7}\Msun) could escape into the intergalactic medium. The rest of the gas will fall back to the central regions, remaining available for additional star formation episodes. The presence of dusty outflows and a clumpy (\ie\,low filling factor) ISM give support to the attenuation-free scenario proposed to explain the high-\z UV-bright compact galaxies such as \RXC. On the other hand, the high densities in the ISM together with the high sta formation rate surface brightness and the amount of returning outflowing mass give support to the feedback-free starburst scenario.} 

\titlerunning{-}
\maketitle

\section{Introduction }
\label{sec:Introduction}

Amongst the major discoveries found by JWST in the distant Universe is the existence of low-metallicity galaxies with enhanced nitrogen (super-solar) abundances at redshifts from \simi5 up to 14.4 (\eg \citealt{Bunker23,Cameron23,Castellano24,Ji24,Schaerer24,Topping24,Curti25b,Naidu26,Stiavelli25,Napolitano25,Zhang26}). These objects, commonly called N-emitter galaxies, usually also present sub-solar C/O ratios, leading to systematically low C/N \citep{Isobe23, Berg26, Ji26}. In addition, many of these N-emitter galaxies show low metallicity and high UV-derived electron densities (\eg \citealt{Topping24,Maiolino_24, Yanagisawa24_He}), high stellar mass and star formation surface densities \citep{Schaerer24,Martinez25,Cameron26}, and/or compact morphologies \citep{Tacchella23,Harikane25}. These properties suggest that nitrogen is enriched in extremely dense and compact starbursts, linking these galaxies to young massive clusters, which are thought to be the progenitors of globular clusters (\citealt{Portegies-Zwart10,Charbonnel23,Marques-Chaves24,Senchyna24,Ji26}).

The scenarios discussed to explain the CNO cycle in these compact galaxies include the enrichment from Wolf-Rayet stellar winds \citep{Senchyna24,Watanabe24, Berg26}, and the presence of super- and very-massive stars (with >100 and >1000\Msun, SMS and VMS, respectively; \citealt{Charbonnel23,Nagele23,Vink25,Ebihara26}). Other scenarios include a wide variety of assumptions, such as pollution from Pop\,III star formation, tidal disruption of stars from encounters with black holes, and ejecta from very massive stars formed through collisions in dense clusters \citep{Cameron23,Senchyna24}. Interestingly, runaway stellar mergers inside these clusters have also been proposed to be one of the main paths to massive black hole seeds \citep{Portegies-Zwart02,Inayoshi20,Trinca23,Rantala24,Partmann25}. In fact, many of these N-enhanced galaxies present broad-line regions (\citealt{ Larson23, Ubler23, Ji24}), in addition to other active galactic nuclei (AGNs) indicators such as [\ion{Ne}{IV}]$\lambda$3426 \citep{Isobe25} or X-ray emission \citep{Napolitano25}, suggesting a connection between nitrogen enhancement, proto-globular clusters, and AGN activity.

Recent JWST observations have shown a complex interstellar medium (ISM) in some N-enhanced galaxies, with a wide range of electronic densities and temperatures derived from rest-frame UV and optical lines \citep{Maiolino_24,Alvarez-Marques25} or the presence of dusty outflows associated with extinction-free, compact UV-bright starburst \citep{Marques-Chaves25}. The presence of a stratified or patchy medium has been discovered in other high-\z galaxies \citep{Ji24,Harikane25,Usui25} as well as local analogues \citep{Berg2021,Mingozzi22,Rickards-Vaught25}. However, it is still unclear whether these properties can explain the extreme physical conditions observed in N-enhanced galaxies and whether they are therefore common to all such objects.

One of the most studied N-enhanced galaxies is RXCJ2248-ID. This multiple-imaged lensed galaxy was first identified by \citet{Boone13} and \citet{Monna14} via the Cluster Lensing And Supernova survey
with Hubble (CLASH, \citealt{Postman12}), while its five images were later confirmed with subsequent spectroscopy data \citep{Schmidt16,Schmidt17,Mainali17}. The detection of its bright and blue UV spectral energy distribution (SED) along with the presence of \ion{C}{IV}]1550 and \ion{O}{III}]1666 emission suggested the presence of a high ionisation field in an extinction-free low-metallicity ISM \citep{Monna14,Mainali17}. Overall, the high intrinsic UV magnitude ($M_\mathrm{UV}$\,\simi-20; \citealt{Topping24}) of \RXC make it the perfect laboratory to investigate the properties and nature of N-enhanced galaxies. In addition, the gravitational magnification ($\mu$\,\simi5\range8) produced by RXC\,J2248.7-4431 (also known as Abell\,S1063), a galaxy cluster at \z{}\,\simi0.35, grants an unprecedented spatial resolution of this galaxy.

Consequently, \RXC became a high-priority target for observations with  JWST, unveiling a physical scenario far more complex than previously known. Specifically, the NIRCam images presented in \citet{Topping24} have revealed that this galaxy is composed of two different compact ($r_\mathrm{eff}$\,\simi22\,pc) clumps separated by \simi200\,pc. This compact size yields extremely large stellar mass and star formation rate (SFR) surface densities ($\Sigma_\mathrm{*}$\simi3.6\EXP{4}\Msun{}\,pc$^{-2}$ and $\Sigma_\mathrm{SFR}$\simi 10.4\EXP{3}\Msyr{}\,kpc$^{-2}$, respectively) that are similar to the densest local stellar clusters \citep{Norris14}. In addition, the NIRSpec Multi Object Spectroscopy (MOS) data obtained with the medium resolution grating (R\,\simi1000) show the presence of multiple narrow high-ionisation lines (\eg \ion{He}{II}, \ion{N}{IV}]) along with weak emission from low-ionisation states ([\ion{O}{III}]/[\ion{O}{II}]\simi200; \citealt{Topping24}). Interestingly, the rest-UV emission lines trace the presence of a highly dense (log(\Ne/\cubiccm{})\eqq4.8\range5.5) nitrogen-enriched (log(N/O)\eqq$-$0.39) ISM, relative to its low metallicity traced by the oxygen abundance (12\,+\,log(O/H)\eqq7.4; \citealt{Topping24}). The spectral resolution of this NIRSpec data revealed the presence of a broad line component in \Ha{} (full width at half maximum, FWHM\,\simi600\kms) and \OIII{5008} (FWHM\,\simi1300\kms) tracing the presence of outflows induced by intense star formation and/or an active galaxy nucleus. 

According to recent models of extreme starbursts in the early Universe \citep{Dekel23,Ferrara23}, galaxies such as \RXC could be in a short-lived bursty outflowing phase. During such as phase, powerful outflows would blow the dusty surrounding medium from the burst region, leaving it as a UV-bright compact source \citep{Fiore23,Ferrara25}. The star formation process would be very effective (\ie with a gas-to-star conversion efficiency close to unity) during a period of time such that supernovae (SNe) and stellar winds would generate outflows and likely intermediate-mass black hole seeds \citep{Li24,Dekel25}.

Recent deep JWST/NIRSpec observations have established RXCJ2248-ID as a crucial benchmark for interpreting chemically enriched multi-phase starbursts in the early Universe. Using deep JWST/NIRSpec spectroscopy, \citealt{Berg26} identified Wolf-Rayet nitrogen stellar features in this galaxy through the detection of broadened \ion{He}{II} and \ion{N}{III} emission. The authors found a strongly elevated and uniform N/O abundance across multiple ionisation zones (log(N/O)=$-$0.391), contrasted by a relative depletion in C/O. The specific abundance pattern indicates that the extreme nitrogen excess in RXCJ2248-ID3 is driven by a short-lived (3\range6\,Myr) phase of nitrogen-rich Wolf-Rayet (\ie WN) stellar winds captured just prior to dilution by subsequent carbon-rich Wolf-Rayet (WC) winds or asymptotic giant branch (AGB) star contributions. Crucially, this interpretation requires the galaxy be observed during a brief evolutionary window following a massive, dense starburst. Consequently, the extreme nitrogen enrichment observed in RXCJ2248-ID, and similar high-redshift systems, likely does not require exotic formation channels but rather represents a transient, ubiquitous phase of bursty star-formation cycles in early low-metallicity compact galaxies. The key properties identified so far in \RXC together with its relatively low-redshift and lensing magnification make this galaxy the best target amongst the N-enhanced galaxies for a detailed  multi-wavelength study involving high spectral resolution JWST integral field spectroscopy and ALMA far-infrared (far-IR) imaging.

In this context, we present a study based on deep high-resolution (R\,\simi2700) NIRSpec/IFU observations of \RXC. These data allowed us to spectrally resolve the different components detected in \Ha{} and \OIIIn{}, and due to the 2D information, we could obtain a better picture of the emission of this source. In addition, comparing the optical and the far-IR ALMA/\OIIIalma emission allowed us to explore lower-ionisation regions of the ISM compared to the rest-UV. The paper is organized as follows. Section~\ref{sec:2.General_observations} presents
the JWST/NIRSpec IFS data used in this work along with ancillary ALMA/\OIIIalma and JWST/NIRCam observations. The analysis we carried out, including the aperture extraction and line fitting, is described in Sect.~\ref{sec:3.Analysis}. In Sect.~\ref{sec:Results}, we derive the main kinematic properties and the electron temperature and density based on the line fitting of the main emission lines, while their interpretation is discussed in Sect.~\ref{sec:Discussion}. Finally, in Sect.~\ref{sec:5.Summary} we summarise the main conclusions
and results of this work. In addition, further details about the ALMA observations and complementary line fitting images can be found in Appendices~\ref{ap1:ALMA} and \ref{ap2:line-fit}, respectively. Throughout this work we assume a flat Lambda cold dark matter ($\Lambda$CDM) cosmology, with $\Omega_\mathrm{m}$\,=\,0.31 and H$_0$\,=\,67.7\,km\,s$^{-1}$\,Mpc$^{-1}$ \citep{PlanckCollaboration18VI}. For this cosmology, 1 arcsec corresponds to 5.79\,kpc at \z{}\eqq6.105, while the luminosity distance is $D_\mathrm{L}$\eqq60.26\,Gpc. In addition, as we have corrected all the fluxes from the \RXCthree image caused by its gravitational magnification ($\mu$\eqq5.3, \citealt{Monna14}), we refer to the galaxy simply as \RXC throughout this paper.

\section{Observations and data calibration}
\label{sec:2.General_observations}

\subsection{NIRSpec IFS observation}
\label{subsec:NIRSpec}

The NIRSpec IFS data were obtained on November 16, 2022, as part of the RIOJA project (PID 1840, PIs: J.\,\'Alvarez-M\'arquez $\&$ T.\,Hashimoto). The observations were taken with a grating+filter pair of G395H\,+\,F290LP, yielding a cube covering the 2.87\,$-$\,5.27\micron wavelength range with a spectral resolution of $R$\,$\sim$\,2700. During the observations, a medium-size cycling dither pattern of four dithers was considered. The total integration time was 5310 seconds using the NRSIRS2RAPID read-out mode.

The raw data were processed with the JWST pipeline version 1.18.0 (\citealt{Bushouse22}) with the context 1355 of the Calibration Reference Data System (CRDS). Following the data reduction process used by previous RIOJA papers (\eg \citealt{Hashimoto23,Mawatari26,Ren25, Sugahara25, Usui25}) and the GTO program (\eg \citealt{Marshall23,Perna23, Ubler23}), we applied some modifications to the pipeline, including (1) the removal of the $1/f$ noise (see \citealt{Bagley23}) and (2) rejection of the bad pixels and cosmic rays through sigma clipping. The data cube was drizzled to a final pixel size of 0.05\arcs{}. Based on the curve of growth from a NIRSpec calibration star (see Sect.~\ref{subsec:int_spec}), we derived a point spread function (PSF) FWHM of \simi0.18 and 0.21\arcs at \OIII{5008} and \Ha wavelengths, respectively. These values, which are in agreement with previous estimations (see \citealt{DEugenio24}), correspond to \simi200\range220\,pc in \RXC.

\subsection{\OIIIalma ALMA observations}
\label{subsec:ALMA}

We retrieved archival ALMA Cycle 2 data (program ID: 2013.1.01052.S, PI: S.\ Madden) taken at Band 8 to search for \OIIIalma. Table~\ref{tab:ALMA-shogen} summarises the details of the observations.  
The Band 8 observations were carried out from May to June 2015 and May to August 2016, yielding a total on-source time of 4.87\,hours. The data were calibrated using a pipeline running on the Common Astronomy Software Application (\textsc{casa}). In particular for the \OIIIalma data obtained in 2015, the raw visibility data were processed using a dedicated pipeline created with \textsc{casa} version 5.6.1 and were carefully inspected for bad data. We imaged the calibrated visibility data using the \textsc{casa} task, \texttt{tclean}, with the natural weighting and a $uv$ taper of $0\farcs 50$ to optimise the sensitivity to the expected source size. The spectral cube was obtained with a frequency step of 39.873\,MHz (25.0\kms). The dust continuum image at rest-frame 90{}\micron was also obtained for line-free channels over 466.22\range468.04\,GHz and 476.25\range478.12\,GHz.
The resulting $1\sigma$ noise levels are 0.70\,mJy/beam (\OIIIalma line), 153\,$\mu$Jy/beam (90\micron continuum) while the beam size is $0\farcs 71$\,$\times$\,$0\farcs62$ (PA\eqq$86.8\degr$), which corresponds to a physical value of \simi750\,pc when considering \RXC redshift and magnification.

We detected a line feature at 477.439\pmm0.063\,GHz with a significance of 4.9$\sigma$ in the integrated intensity, strongly suggesting the \OIIIalma line at \z{}\eqq6.107\pmm0.001. This is consistent with the optical NIRSpec redshift. Figure~\ref{fig:regions_spectra} shows the \OIIIalma image contours, and the peak position obtained by a Gaussian fit is ($\alpha_\mathrm{IRCS}$, $\delta_\mathrm{IRCS}$) = ($22^{\rm h} 48^{\rm m} 45\fs 808$, $-44\degr 32\arcmin 14\farcs 749$). The spectrum obtained at the peak position is shown in Figure~\ref{fig:alma_spec}. The line FWHM is 174\pmm43\kms{}, while the observed integrated flux and intrinsic (\ie magnification corrected) luminosity are $F_\mathrm{[\ion{O}{III}]\,88\mu m}$\eqq0.49\pmm0.10\,Jy\kms and $L_\mathrm{[\ion{O}{III}]\,88\mu m} = (1.7 \pm 0.3) \times 10^8$\Lsun, respectively (see Table~\ref{table:alma_results}).

We did not detect the rest-frame 90$\mu$m continuum emission, resulting in $2\sigma$ upper limits of $S_\mathrm{90\mu m}$\,<\, 0.31\,mJy (Table~\ref{table:alma_results}).  If we assume $T_{\rm dust} = 50$~K (80~K) and $\beta_{\rm dust} = 2$ as the dust temperature and emissivity index, the intrinsic total infrared luminosity is $L_{\rm TIR}$\,<\,7.5\EXP{10}\Lsun (3.8\EXP{11}\Lsun).

\begin{table}
 \caption{ALMA results for \RXCthree.\label{table:alma_results}}
    \centering
    \begin{tabular}{lc}
    \hline
    \hline
    Parameter$\dag$ & Value  \\
    \hline
    $F_\mathrm{[\ion{O}{III}]\,88\mu m}$ & 0.49\pmm0.10\,Jy $\mu^{-1}$\kms  \\
                       & (14.8\pmm1.8)\EXP{-19}\ergscm  \\
    $S_{90\mu m}$           & <\,0.31\,$\mu^{-1}$\,mJy  (2$\sigma$)  \\
    $L_\mathrm{[\ion{O}{III}]\,88\mu m}$ & (1.7\pmm0.1)\EXP{8}\Lsun \\
    $L_\mathrm{TIR}$      & <7.5 \EXP{10}\Lsun (2$\sigma$) \tablefootmark{$\sharp$}\\
    \hline
    \end{tabular}
\tablefoot{
\tablefoottext{$\dag$}{Intrinsic values were derived considering a lensing magnification $\mu$\eqq5.3.}
\tablefoottext{$\sharp$}{The dust temperature and emissivity index are assumed to be $T_{\rm dust} = 50$~K and $\beta_{\rm dust} = 2$, respectively. If $T_{\rm dust} = 80$~K is assumed, the limit becomes $<$3.8\EXP{11}\Lsun.}
}
\end{table}

\subsection{Complementary NIRCam images}
\label{subsec:NIRCam}

We made use of the available RIOJA NIRCam images for \RXCthree to align the NIRSpec data with ALMA and to depict the morphological distribution of UV and optical light. Concretely, we used the F115W and F444W filters, which cover the rest-frame UV and the \Ha emission, respectively. These images were taken on September 24, 2022, with a total exposure time of 1245\,s using the SHALLOW4 readout mode.

The NIRCam data were calibrated using the JWST pipeline v1.12.3 \citep{Bushouse22} with the context 1130 of the CRDS. We also applied additional steps (see \citealt{Hashimoto23,Sugahara25} for further details), such as removal of snowballs and wisps as
described in \citep{Bagley23} and background
homogenization as applied in \citet{Perez-Gonzalez23b}, and we incorporated the improvements presented in \citet{Perez-Gonzalez25} and \citet{Ostlin25}. The final images were drizzled to a pixel scale of 0.03\arcs/pixel.

\subsection{Astrometric alignment}
\label{subsec:astrometry}

Before analysing the data, we aligned the JWST and ALMA data. First, we used the available Gaia DR3 \citep{GaiaCollaboration23} stars within the NIRCam field of view to align these images, yielding an uncertainty smaller than 1 pixel (\ie 30\,mas) in the final absolute positioning. Then, following \citet{Mawatari26} and \citet{Usui25}, we cross-correlated a NIRSpec/\Ha map, created by integrating 0.02\micron around the emission line, with the NIRCam F444W image, which contains its emission. The ALMA astrometry is referenced to quasars, yielding an absolute uncertainty of a few milliarcseconds. We therefore assumed that after our NIRCam+NIRSpec alignment, we have both JWST and ALMA datasets in the same reference frame. 

We note in Fig.~\ref{fig:regions_spectra} that the \OIIIalma peak is centred in \RXC, while the NIRSpec/\OIII{5008} emission is offset by \simi0.1\arcsec, being spatially coincident with the UV-peak observed in NIRCam/F115W. Considering the uncertainties in the alignment and the NIRCam PSF, it seems that the \OIIIalma emission is much more centred and produced homogenously from the different clumps in the galaxy, while most of the UV and \OIII{5008} emission is located in the right clump.

\section{Data processing and analysis}
\label{sec:3.Analysis}

\subsection{Aperture extraction and line fluxes}
\label{subsec:int_spec}

\begin{figure*}
\centering
   \includegraphics[width=\linewidth]{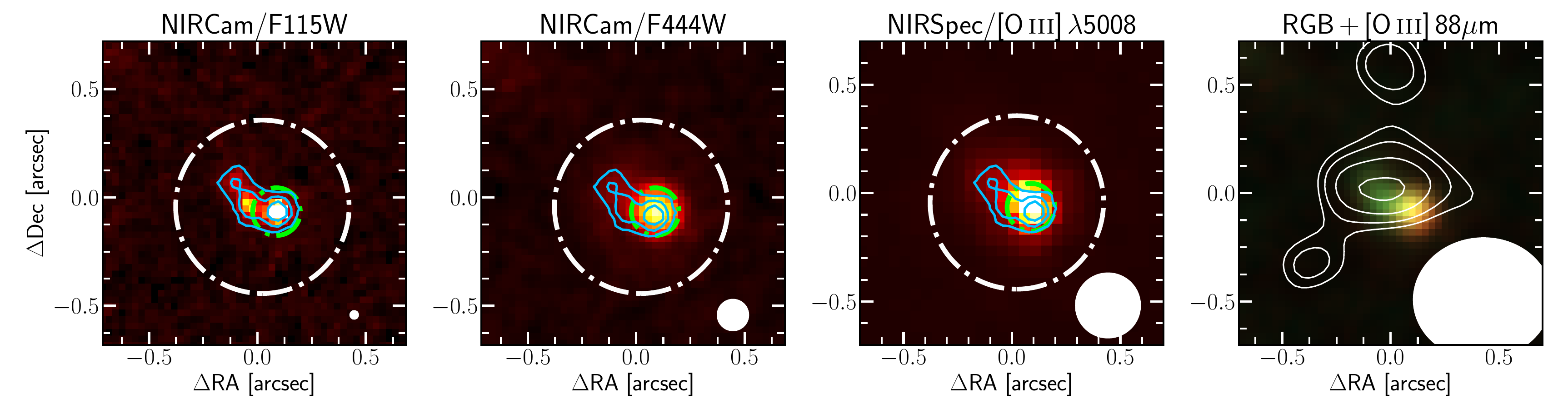}
      \caption{NIRCam F115W and F444W, NIRSpec/\OIII{5008} and RGB (F115W/F200W/F444W) cutouts for \RXCthree. Blue contours show the F115W emission. White and green circles represent the \total and \nuclear apertures used to extract the integrated spectra, respectively. White circles (ellipse) in the lower right corners display the FWHM (beam size) for the NIRCam/NIRSpec (ALMA) data. White contours in the RGB image represent the \OIIIalma emission integrated over 477.27\range477.75\,GHz. The contours are drawn at 2, 3, 4 and 5$\sigma$, where $\sigma = 70$\,mJy\kms{}\,beam$^{-1}$.}
         \label{fig:regions_spectra}
\end{figure*}

\begin{figure*}
\centering
   \includegraphics[width=\linewidth]{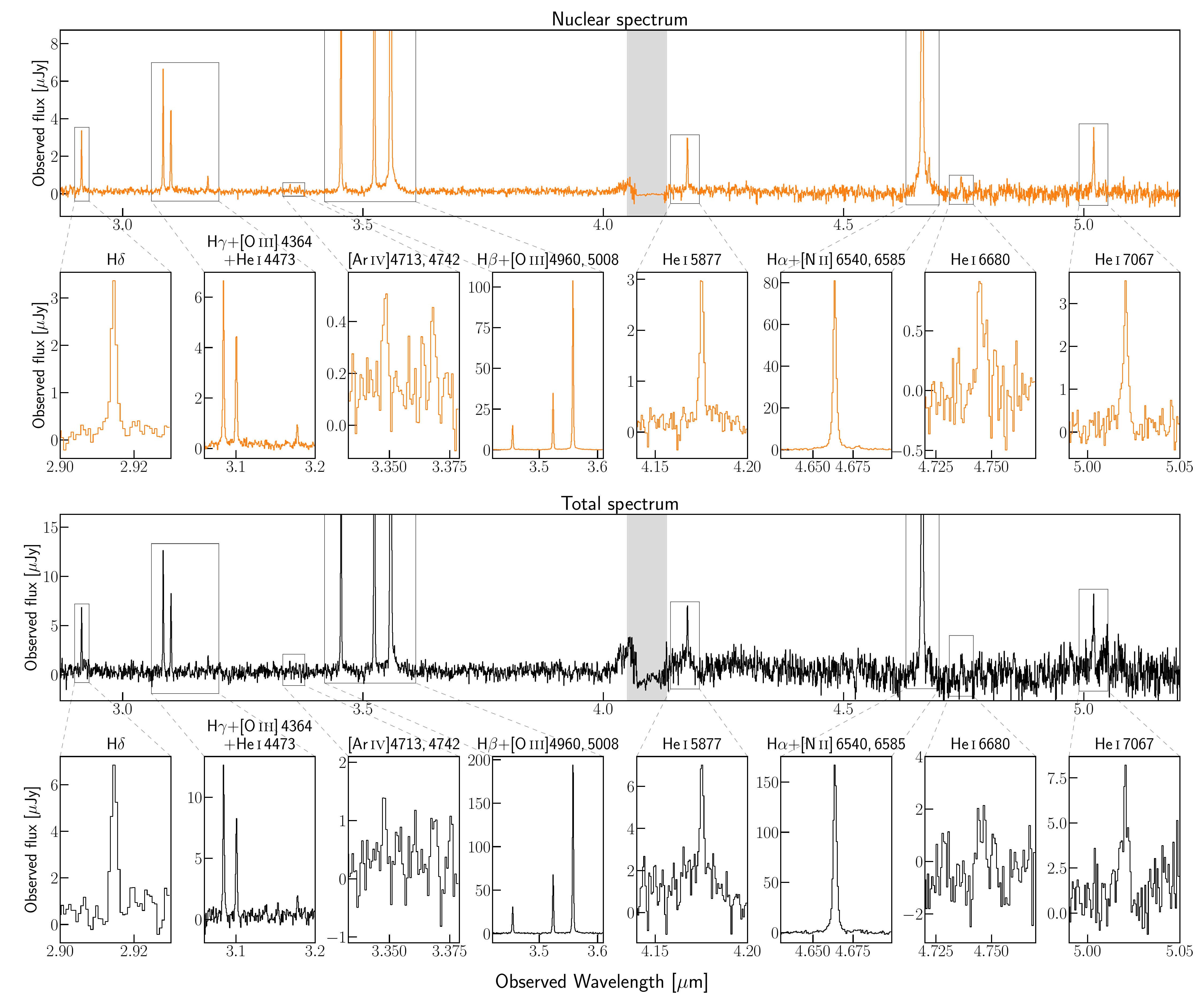}
      \caption{Observed spectra for the \nuclear and \total apertures. The zoom-in sub-panels show the main emission lines detected in this work. Grey areas mark the spectral regions affected by the detector gap in NIRSpec.}
         \label{fig:lines}
\end{figure*}

We analysed the optical nebular emission of \RXC using two different circular apertures: a large one ($r$\eqq0.4\arcs; \total) covering the entire galaxy and a smaller aperture centred on the brightest clump ($r$\eqq0.1\arcs; \nuclear). These apertures were defined to enclose all the NIRSpec emission and optimise the signal-to-noise (S/N) of the spectra at the centre of the brightest clump, respectively. Figure~\ref{fig:regions_spectra} shows the circular apertures used to extract the spectra of these regions. In addition, we derived an average background spectrum considering an annular-aperture (0.8\arcs{}$<$\,$r$\,$<$\,1.5\arcs) that is used to clean the \total and \nuclear spectra of residual features introduced by the reduction process and background emission.

The background-clean spectra reveal several emission lines across the NIRSpec wavelength range. Concretely, we detected four Balmer lines (\Ha, \Hb, \Hg, \Hd), three \OIIIn{} lines (\OIII{4364,\,4960,\,5008}), and four \HeIn lines (\HeI{4473,\,5877,\,6680,\,7067}). Additionally, we detected the \ArIV{4713,\,4742} and \NII{6540,\,6585} doublets in the \nuclear spectra due to the higher S/N. In Figure~\ref{fig:lines} we present the main emission line for the \total and \nuclear apertures.

Emission line profiles were modelled using the \texttt{astropy.modeling} framework, employing the \texttt{TRFLSQFitter} (trust region reflective algorithm) to perform a non-linear least-squares minimisation. To ensure robust parameter estimation and realistic uncertainty quantification, we implemented a Monte Carlo approach with 1000 iterations, adding random noise characterised by a standard deviation equal to the root mean square of the spectra. This root mean square level was measured around each fitted line using spectral windows of 0.2\micron, masking the emission line contribution. The resolving power in our NIRSpec data varies with the wavelength, ranging between $R$\,\simi2000 and 3600.\footnote{More info: \href{https://jwst-docs.stsci.edu/jwst-near-infrared-spectrograph/nirspec-instrumentation/nirspec-dispersers-and-filters\#gsc.tab=0}{JWST User Documentation}} We therefore took into account this variation when measuring the intrinsic width of each line component by correcting for the instrumental FWHM at each wavelength (\ie 128, 124, and 93\kms for \Hb, \OIII{5008}, and \Ha wavelengths, respectively).

\begin{figure*}
 \sidecaption
   \includegraphics[width=12cm]{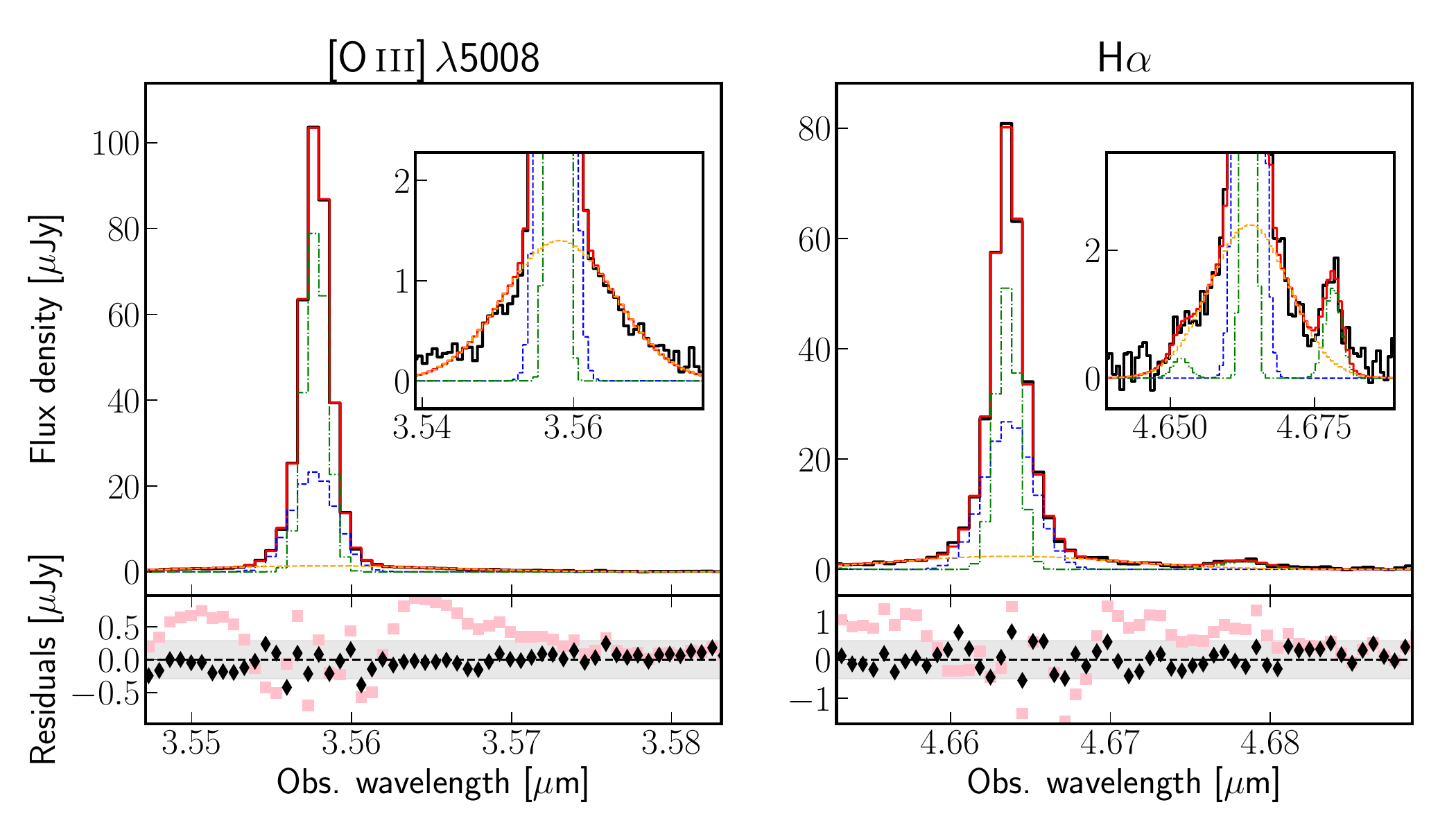}
      \caption{Multiple Gaussian analysis performed in the observed \OIII{5008} and \Ha\,+\,\NII{6550,\,6585} lines extracted from the \nuclear aperture (see Sect.~\ref{subsec:Kin}). Solid black and red lines represent the observed line and its best-fit model, while the narrow, broad and very broad components are shown in doted-dashed green, and dashed blue and orange lines, respectively. In the lower panels, black diamonds show the residuals between the best-fit model and the observed line profiles, where the grey shadow area represents the 3$\sigma$ noise derived during the fitting procedure (see Sect.~\ref{subsec:int_spec}). For comparison, pink squares represent the residuals assuming 2 Gaussian profiles for the \OIII{5008} and \Ha lines. Each panel contains a zoom-in inset to display the contribution of the broad and very broad components to the line profiles.}
         \label{fig:Broad_fitfig}
\end{figure*}

We followed different approaches during the fitting procedure. First, we adopted single-Gaussian profiles for all the emission lines detected. Then, we assumed multiple-Gaussian profiles for those lines showing broad emission and large residuals using the single-Gaussian approach (\ie \Hb, \OIII{4960}, \OIII{5008}, and \Ha; see Sect.~\ref{subsec:Kin} and Fig.~\ref{fig:Broad_fitfig}). For \Ha we adopted three independent Gaussian profiles, while we included the presence of the narrow \NII{6550,\,6585} doublet, fixing their ratio to 1:3. We simultaneously fitted \Hb with the same three Gaussian profiles, fixing the FWHM and velocity offset to match the \Ha ones. We applied a similar approach to \OIII{4960} and \OIII{5008} lines, which were simultaneously fitted with three Gaussians, constraining the same FWHM and a 3:1 ratio in amplitude between the corresponding components of both lines. We compared these three-component fits with an alternative model using only two components per line, where the Akaike information criterion (AIC; \citealt{Akaike74})\footnote{AIC\,=\,2$k-$2log($\hat{L}$), where \textit{k} is the number of free parameters in the model and $\hat{L}$ is the maximised value of the likelihood function. This index balances goodness of fit against the model complexity.} shows that the three-Gaussian approach offers, in general, a better fit ($\Delta$AIC\eqq AIC$_\mathrm{2G}-$AIC$_\mathrm{3G}$\,$>$10). For completeness, we tested whether the two- or three-component fits can be extended to \Hd, \Hg, and \OIII{4364}. The AIC analysis revealed that although we cannot discard the presence of a broad component in \Hg and \OIII{4364}, these lines are better described by single-Gaussian models. In the following sections, we refer to the three components found in \OIII{4960}+\OIII{5008} and \Ha as `systemic', `broad', and `very broad'. In general, we observed that a three-component fit is also favoured when using the \total aperture, suggesting that the \nuclear contribution dominates the total emission. Only for the \Ha and \Hb lines from the 'total' spectrum is a two-Gaussian fit preferred, due to the lower S/N in the larger aperture. Figure~\ref{fig:Broad_fitfig} shows the best-fit models for \OIII{5008} and \Ha lines, and the distinct components can be observed. For illustrative purposes,  the lower panels show the residuals when considering three and two Gaussian profiles, respectively. Complementary plots showing the simultaneous \OIIIn fitting. The best-fit model for \Hb and the single-Gaussian fit for the \ArIVn doublet are presented in Appendix~\ref{ap2:line-fit}.

All the derived fluxes were corrected for aperture losses using a calibration star as follows. First, we reduced the NIRSpec calibration star TYC 4433-1800-1 (PID 1128) using the same pipeline applied to the science data (see Sect~\ref{subsec:NIRSpec}). Then, we created pseudo-continuum images by integrating every 50 spectral channels (\ie \simi300\AA) to sample the wavelength-dependent evolution of the PSF. Finally, for each pseudo-continuum image, we computed the curve of growth and derived the corresponding aperture correction at each radius. Once the wavelength-dependent evolution of the aperture correction was determined, we interpolated the value for each emission line wavelength and applied the corresponding aperture correction. These corrections range from \Hd to \HeI{7067} within 2.1\range2.7 and 1.1\range1.2 for the `nuclear' and `total' apertures, respectively. Table~\ref{tab:fluxes} presents the de-magnified aperture-corrected fluxes for all the emission lines detected in each aperture.

\begin{table*}[!ht]
\caption{\RXC line fluxes.}
\centering
\begin{tabular}{l|c|c|c|c|c|c}
 Line & $F_\mathrm{narrow}^\mathrm{tot}$ & $F_\mathrm{broad}^\mathrm{tot}$ & $F_\mathrm{very-broad}^\mathrm{tot}$ & $F_\mathrm{narrow}^\mathrm{nuc}$ & $F_\mathrm{broad}^\mathrm{nuc}$ & $F_\mathrm{very-broad}^\mathrm{nuc}$ \\
\hline
\Hd & 9.8\pmm0.4 & - & - & 7.4\pmm0.2 & - & - \\
\Hg & 16.5\pmm0.4 & - & - & 14.7\pmm0.2 & - & - \\
\OIII{4364} & 11.6\pmm0.4 & - & - & 12.2\pmm0.2 & - & - \\
\HeI{4473} & 2.3\pmm0.4 & - & - & 1.9\pmm0.2 & - & - \\
\ArIV{4713}$^{\dagger}$ & - & - & - & 0.96\pmm0.13 & - & - \\
\ArIV{4742} & - & - & - & 0.71\pmm0.12 & - & - \\
\Hb & 33\pmm1 & 5\pmm1 & - & 19\pmm1 & 13\pmm1 & 3.6\pmm0.5 \\
\OIII{4960} & 51\pmm5 & 27\pmm5 & 6.9\pmm0.9 & 46\pmm1 & 25\pmm1 & 8.0\pmm1.1 \\
\OIII{5008} & 149\pmm16 & 80\pmm13 & 20\pmm2 & 133\pmm2 & 74\pmm2 & 23\pmm1 \\
\HeI{5877} & 8.9\pmm1.0 & - & - & 5.0\pmm0.2 & - & - \\
\Ha & 95\pmm3 & 48\pmm2 & - & 51\pmm4 & 56\pmm5 & 21\pmm1 \\
\NII{6585} & 3.8\pmm0.5 & - & - & 4.9\pmm0.1 & - & - \\
\HeI{6680} & 1.3\pmm0.8 & - & - & 1.8\pmm0.2 & - & - \\
\HeI{7067} & 8.0\pmm0.4 & - & - & 5.8\pmm0.1 & - & - \\
\hline
\OIIIalma & 14.8\pmm1.8 & - & - & -  & - & -

\label{tab:fluxes}
\end{tabular}
\tablefoot{De-magnified ($\mu$\eqq5.3) emission line fluxes (in \EXP{-19}\ergscm) derived from the \total and \nuclear apertures, respectively. Fluxes have been corrected from aperture loses (see Sect.~\ref{subsec:int_spec}). Narrow, broad, and very broad component values represent the three different kinematic components derived in Sect.~\ref{subsec:Kin}. These line fluxes have not been corrected from dust reddening. \ArIV{4713}$^\dagger$ displays the observed blended emission for \ArIV{4713}\,+\,\HeI{4714}. The fluxes corrected by the \HeI{4714} contribution assuming the `primordial' and `overabundant' metallicity scenario are (0.61\pmm0.14) and (0.58\pmm0.14)\EXP{-19}\ergscm, respectively (see Sect.~\ref{subsec:deblending}).}
\end{table*}

\subsection{Deblending of \ArIV{4713}\,+\,He\,I\,4714 lines }
\label{subsec:deblending}

In the subsequent analysis, the electron density of the ionised gas is derived using the ratio of the \ArIV{4713}/\ArIV{4742} doublet. Since the \ArIV{4713} line is blended at the NIRSpec R2700 resolution with the \HeI{4714} and \NeIV{4716} lines, we first had to derive the contribution of these lines to the total measured flux. We first assumed that the contribution of the \NeIV{4716} doublet is negligible. In star-forming galaxies (SFGs), the [\ion{Ne}{IV}] emission is expected to be much lower than \HeIn due to its larger ionisation potential (\ie \simi63.5\,eV). On the other hand, the \NeIV{4727} doublet, which is typically \simi35$\%$ brighter than \NeIV{4716} in narrow-line regions (\citealt{Binette24}), is not detected. Therefore, we only considered the \HeI{4714} contribution during the deblending process.

We computed the \HeI{4714} flux based on its ratio with the other optical \HeIn lines detected in the \nuclear spectrum. For this analysis, we discarded the \HeI{7067} line, as its strong sensitivity to collisional excitation and radiative transfer effects (\eg fluorescence) might boost its intensity and lead to spurious results \citep{Izotov97,Benjamin02,Monreal-Ibero17}. However, it should be noted that the other optical \HeI\ lines can also be moderately affected by radiative transfer and collisional effects \citep[see e.g.][]{Berg26a}, which introduce a minor systematic uncertainty. We therefore derived the theoretical line ratios \HeI{4714}/\HeI{4473}, \HeI{4714}/\HeI{5877}, and \HeI{4714}/\HeI{6680} using \texttt{Pyneb} \citep{Luridiana15}, assuming the negligible dust extinction derived for the narrow lines (see Sect.~\ref{subsec:extinction}) and following the results of \citet{Yanagisawa24_He}. The authors of that work presented two different scenarios to explain the strong helium and nitrogen emission in \RXC. Their first scenario proposes an overabundance of helium linked to the CNO cycle. The second one, however, suggests an ionised ISM with a primordial He abundance and high electron densities that would cause an excess of collisional excitation and therefore an increase of HeI emission relative to hydrogen lines. The authors derived electron densities and temperatures of \Ne{}\eqq1148$^{+1143}_{-407}$\cubiccm and \Te{}\eqq28012$^{+1428}_{-6060}$\K for the `primordial' HeI abundance scenario and of \Ne{}\eqq9$^{+38}_{-7}$\cubiccm and \Te{}\eqq23241$^{+1711}_{-1686}$\K for the `overabundant' scenario.

Applying the \texttt{Pyneb} theoretical He\,I ratios for these \Te and \Ne pairs to the He\,I lines measured in the \nuclear aperture, we obtained average \HeI{4714} fluxes of (3.5\pmm0.3) and (3.8\pmm0.5)\EXP{-20}\ergscm for the `primordial' and `overabundant' scenarios, respectively. These fluxes (and their associated uncertainties) are defined as the average (and the standard deviation) of the fluxes derived using the different line ratios. The low standard deviation for the fluxes confirms that the values derived from the individual line ratios are consistent within their uncertainties. These flux values represent \simi35\range40$\%$ of the total flux measured for the blended \ArIV{4713}\,+\,\HeI{4714} emission.

\section{Results}
\label{sec:Results}

\subsection{Kinematic substructure in the ionised ISM}
\label{subsec:Kin}

\begin{table*}[!ht]
\caption{Intrinsic kinematic properties for the ionised gas from the multi-Gaussian line fitting.}
\centering
\begin{tabular}{l|c|c|c|c|c|c|}
  & FWHM$_\mathrm{sys}$ & FWHM$_\mathrm{b}$ & $\Delta V_\mathrm{b}$ &  FWHM$_\mathrm{vb}$ & $\Delta V_\mathrm{vb}$ \\
    & [\kms]  & [\kms]  & [\kms]  & [\kms] & [\kms] \\
\hline
\OIII{5008} & 73\pmm1 & 248\pmm4 & -10\pmm1 & 1456\pmm87 & 30\pmm9 \\
\Ha & 60\pmm6 & 213\pmm12 & 9\pmm1 & 1005\pmm96 & 11\pmm11
\label{tab:Broad_fit}
\end{tabular}
\tablefoot{`sys',`b' and `vb' suffix stand for the narrow (\ie systemic velocity), the broad and very broad components, respectively. $\Delta V$ values correspond to the velocity offset between the (very-)broad and the narrow components.}
\end{table*}

The multi-Gaussian analysis carried out in Sect.~\ref{subsec:int_spec} revealed the presence of three distinct kinematics components in the ISM, visible in \OIII{5008} and \Ha{} in the \nuclear spectrum due to their high S/N. In general, we find agreement in the kinematics of the systemic component in the emission lines, with an instrumental FWHM of\,\simi60\range73\kms (see Table~\ref{tab:Broad_fit}). For the secondary component in \OIII{5008} and \Ha{}, we obtained broader FWHMs (\simi213\range248\kms) without significant blue- or red-shifts with respect to the systemic component and with a velocity offset ($\Delta V_\mathrm{b}$) smaller than the size of a spectral channel (\simi30\range60\kms). For the very broad components, we measured an FWHM of\eqq1456\pmm87 and 1005\pmm96\kms in \OIII{5008} and \Ha, respectively, with no significant velocity offsets. This third kinematic component corresponds to the broad component identified in \citet{Topping24} (FWHM\,\simi800\range1400\kms) using NIRSpec/MSA data with a lower spectral resolution (R\,\simi1000). The difference in the FWHM between \Ha and \OIII{5007} for this component was also observed in \citet{Topping24}, and it might arise because \OIIIn traces (in general) lower-density highly ionised gas, which can be easily accelerated to higher velocities in the presence of an AGN \citep{Harrison16, Kang17, Ubler23}.

While \Hd, \Hg, and \OIII{4364} are well described by single-Gaussian fits, the resulting line widths (\ie \simi150\kms) are noticeably larger than those measured for the narrow component in \Ha{} and \OIII{5008}. Although their line profiles might hint at the presence of a broader underlying component (Fig.~\ref{fig:lines}), their lower S/N does not allow for a statistically meaningful multi-Gaussian decomposition.

Based on the results from the low- and high-resolution NIRSpec data, our analysis revealed a scenario where the ISM presents three distinct kinematics components. A dynamically cold gas component with an intrinsic FWHM\,\simi65\kms and two additional gas components characterised by broader (FWHM\,\simi250\kms and \simi1000\kms) line profiles potentially linked to outflowing gas. In Sect.~\ref{sec:Discussion} we discuss the possible physical scenarios powering these outflows and the potential explanations for having such kinematically complex structures in a very compact source ($<$\,22\,pc).

\subsection{Dynamical versus stellar mass}
\label{subsec:dynamicalmass}

An estimate of the dynamical mass of \RXC can be derived from the line width of the systematic component (i.e. the narrow line component with FWHM\eqq65\kms) and the size of the UV-bright stellar population (i.e. effective radius $<$22\,pc, \citealt{Topping24}). Assuming that the system is virilized (see \eg \citealt{Bellocchi13}), we derived a \Mdyn\simi(1\range3)\EXP{7}\Msun, which is four to ten times smaller than its SED-derived stellar mass (\ie log(\Mstar/\Msun)\eqq8.05, \citealt{Topping24}). This discrepancy could be due to the fact that we assumed the ionised gas emitting region has the same size as the UV-bright compact stellar population. If the size of the ionised nebula were larger, \ie 100\range200\,pc, the dynamical mass would be larger by factors of five to ten, in closer agreement with the stellar value. However, the angular resolution of the NIRSpec data did not allow us to resolve the extent of the ionised gas below the PSF size (\ie \simi200\,pc). 

\subsection{Dust distribution in the ionised ISM}
\label{subsec:extinction}

The detection of \Ha and \Hb in all the three kinematic components allowed us to use the Balmer decrement to estimate the extinction affecting them. This dust extinction plays a critical role when comparing optical lines with infrared ones (\eg \OIII{5008}/\OIIIalma), as the latter are much less sensitive to dust attenuation. 

We compared the measured \Ha/\Hb for each component in the nuclear spectra with the case B recombination values derived from \texttt{Pyneb}, adopting the \citet{Cardelli89} extinction law. Concretely, we assumed an intrinsic ratio of \Ha/\Hb{}\eqq2.69 for the electron temperatures and densities derived in this work (\Te{}\eqq34800\,K and log(\Ne{}[\cubiccm{}])\,\simi3; see Sect~\ref{subsec:temp_OIII} and \ref{subsec:density}). This ratio remains almost constant for the higher densities (log(\Ne{}[\cubiccm{}])\,\simi5) found in \citet{Topping24} using UV lines.

For the narrow component, we derived a \Ha/\Hb{}\eqq2.7\pmm0.3, consistent with the negligible extinction found in previous works using NIRSpec/MSA (\citealt{Topping24}). In addition, the ratios \Hb/\Hg and \Hb/\Hd are also compatible with no extinction. We therefore assumed \Av{}\eqq0 for the narrow component in the subsequent sections. In contrast, we found a large Balmer decrement for the broad and very broad components with \Ha/\Hb{}\eqq4.3\pmm0.5 and 6.0\pmm0.9, corresponding to \Av{}\eqq1.5\pmm0.4 and 2.5\pmm0.5\,mag, respectively. These extinction values were calculated assuming the \citet{Cardelli89} attenuation law. This choice is supported by \citet{Reddy20}, who showed that the attenuation curve derived from more than 500 SFGs at \z\simi2 is consistent with the Milky Way curve. Recent JWST studies have also shown that the \citet{Cardelli89} and \citet{Reddy20} attenuation curves give consistent Balmer decrements at \z\simi1\range3 \citep{Reddy23}. However, if we used the steeper extinction curve from \citet{Sanders25} derived for a SFG at \z{}\eqq4.4, our extinction estimates from the H$\alpha$/H$\beta$ ratio would be systematically lower by a factor of about 1.5 in $A_V$ due to its steeper slope at $\lambda>5000$\,\AA. Although further high-redshift studies are still needed to constrain the shape of extinction curves at $z>4$, the potential differences in $A_V$ do not alter the main conclusions of this work. The extinction values derived are in agreement with previous studies showing a high dust attenuation associated with broad components in high-\z SFGs \citep{Lamperti24,Rodriguez-delPino24,Parlanti25}. Further analysis on the \Ha/\Hb spatial distribution and dust scenarios will be presented in Hamada et al. (in prep).

Interestingly, we find that the narrow and the broader components in \RXC exhibit different extinction values despite being observed within an extremely compact region (r\eqq0.1\arcs for \nuclear aperture). The narrow component appears to be in agreement with a dust-free high temperature case B recombination, while both broader components are heavily obscured. This differential extinction was carefully taken into account when determining the optical-to-far-IR \OIII{5008}/\OIIIalma line ratio, which we use in subsequent sections to estimate the electron density, while its physical implications are discussed in Sect.~\ref{sec:Discussion}.

\subsection{Electron temperature}
\label{subsec:temp_OIII}

We computed the electron temperature of the high ionised ISM in \RXC using the \OIII{4364}/\OIII{5008} ratio, which is highly sensitive to temperature variations. As presented in Sect.~\ref{subsec:int_spec} and \ref{subsec:Kin}, we identified three distinct kinematic components in \OIII{5008} and one in \OIII{4364}. The FWHM derived from the single-Gaussian fit in \OIII{4364} is \simi150\kms, representing an intermediate value between the narrow and broad component detected in \OIII{5008} (see Table~\ref{tab:Broad_fit}). Based on this FWHM value, we assume that the \OIII{4364} flux derived from the single Gaussian contains some broad emission, though we are not able to kinematically resolve it due to its lower S/N (see Sect.~\ref{subsec:int_spec}). Therefore, we considered both the narrow and broad components from \OIII{5008} when deriving the \OIII{4364}/\OIII{5008} ratio. Based on the \Av value for the broad component derived in Sect.~\ref{subsec:extinction}, \OIII{4364} is \simi1.5 times more affected than \OIII{5008} by dust obscuration. Thus, we took this value into account to correct the broad components for extinction, considering the same broad-to-narrow ratio in both lines while adopting \Av{}\eqq0 for the narrow components (see Sect.~\ref{subsec:extinction}). Based on the flux values from Table~\ref{tab:fluxes}, we derived a log(\OIII{4364}/\OIII{5008})\eqq$-$1.24\pmm0.05 and $-$1.17\pmm0.02 for the \total and \nuclear apertures, respectively. We note that this approach assumes kinematic similarity of the \OIIIn profiles and that under extreme low broad-contribution cases (0\range20\% of flux in the broad component), the resulting deviations remain moderate, with log(\OIII{4364}/\OIII{5008}) varying between $-$1.6 and $-$1.3. Higher broad-component fractions are not favoured by the \OIII{4363} line fitting (see Sect.~\ref{subsec:int_spec}).

We then compared these line ratios with the \texttt{Pyneb} theoretical ones to estimate the electron temperature. Concretely, we performed a Monte Carlo simulation to sample the line ratio and \Ne parameter space and derived the \Te for the log(\Ne{}[\cubiccm{}])\,<\,4.5 range. In this regime, the \OIII{4364}/\OIII{5008} ratio is primarily sensitive to temperature and only weakly dependent on density, whereas for densities approaching the critical density of \OIII{5008} (\ie\ log($n_\mathrm{c}$[\cubiccm{}]),\simi5.8), the ratio becomes increasingly density dependent. We obtained an average value of \Te{}\eqq30200\pmm3200\K and \Te{}\eqq34800\pmm1000\,K for the \total and \nuclear regions, respectively.

These values are slightly higher than those derived by \citet{Topping24} using the same lines with NIRSpec/MSA (\ie\ \Te{}\eqq24600\pmm2600\K). In that work, the authors adopted log(\Ne{}[\cubiccm{}])\eqq5 based on UV density diagnostics. However, the highest densities were inferred from \ion{N}{IV}], which traces gas with a substantially higher ionisation potential than \OIIIn. In contrast, diagnostics more closely associated with the \OIIIn-emitting region, such as \ArIVn and \ion{C}{III}], yielded lower densities, log(\Ne{}[\cubiccm{}])\simi4\range5 \citep{Topping24,Berg26}. Over this density range, the inferred \Te\ changes by less than the uncertainty associated with the measured \OIII{4364}/\OIII{5008} ratio. Moreover, the density diagnostics themselves depend on the assumed electron temperature such that \Te\ and \Ne\ cannot be determined fully independently. Given that the uncertainty in \Te\ dominates the error budget, we adopted a fiducial density range and derived \Te\ within it. For reference, assuming log(\Ne{}[\cubiccm{}])\eqq5 would decrease the inferred temperature to \Te{}\eqq23200\pmm2700\,K. The combination of the different density assumptions and the de-reddening applied in this work explains the lower \Te\ reported by \citet{Topping24} and \citet{Berg26}.

Our \Te values are also slightly larger than those previously derived for \HeIn (\ie\simi23000\range28000\K, \citealt{Yanagisawa24_He}). Obtaining \OIIIn electron temperatures higher than those from \HeIn has been recently discovered to be common in \ion{H}{II} regions from SFGs and planetary nebulae \citep{Mendez-Delgado25}. According to that work, this \Te discrepancy could arise from a deviation from Case B where photons are absorbed by \HIn rather than \HeIn and/or there is a generalised ionising photon escape or from temperature inhomogeneities induced by a clumpy substructure in metallicity. 

Based on our \Te{}[O$^{++}$] value, we could estimate the electron temperature expected in the single-ionised oxygen region that can be used as a proxy for \HIn regions, as they present similar ionisation potentials (\simi 13.6\,eV). Using the expression \Te{}[O$^{+}$]\eqq0.7\,$\times$\,\Te{}[O$^{++}]$\,+\,3000 from \citet{Campbell86}, we derived a \Te{}[O$^{+}$]\eqq24100\pmm2200\,K for the \total aperture. This value is in agreement with those obtained when applying recent temperature-temperature relations calibrated with \z$>$\,2 galaxies observed with NIRSpec. Concretely, we derived a \Te{}[O$^{+}$]\eqq21100\pmm1000\,K and 22100\pmm6000\,K using the \citet{Cataldi25} and \citet{Chakraborty25} relations, respectively. These values confirm that the electron temperature for even the low ionisation ISM is very high, as previously derived from the \HeIn lines by \citet{Yanagisawa24_He} (see Table~\ref{tab:ne_results} for details).

The detection of these high \OIIIn electron temperatures would indicate the presence of extreme physical conditions. In fact, very broad (FWHM\,$\gtrsim$\,1000\kms) kinematic components have been observed in AGN-driven outflow in similar N-enhanced galaxies \citep{Larson23,Ubler23}. In addition, a very low metallicity (12\,+\,log(O/H)\eqq7.4) has been derived based on the optical lines using NIRSpec/MSA data \citep{Topping24}. We suggest that the high \Te values found in this work result from either a hard ionising radiation field produced by an AGN or a low metallicity environment, which would reduce the cooling efficiency, or a combination of both.

\subsection{Electron density}
\label{subsec:density}

\begin{table*}[!ht]
\caption{Electron densities and temperatures of \RXC traced by UV to far-IR emission lines. }
\centering
\begin{tabular}{l|c|c|c|c}
 Line Tracer & IP & \Ne &  \Te & Reference \\
  & (eV) & (cm$^{-3}$) & (K) & \\
\hline
\ArIV{4713,\,4742}& 40.7 & 6700\oddpm{9351}{3800} & - & This work \\
\OIII{4364,\,5008,\,88$\mu$m} & 35.1 & 500\oddpm{300}{180} & 30200\pmm3200 & This work \\
\OIII{4364,\,5008} & 35.1 & -- & 24600\pmm2600 & \citealt{Topping24} \\
He\,I\,-\,primordial & 24.6 & 1148$^{+1143}_{-407}$ & 28012$^{+1428}_{-6060}$ & \citealt{Yanagisawa24_He} \\
\ion{O}{III}]\,$\lambda$1660,\,1666& 35.1 & -- & 23300\pmm4100 & \citealt{Topping24} \\
\ion{N}{IV}]1483,\,1486 & 47.4 & 3.1$^{+0.5}_{-0.4} \times 10^5$ & 24600\pmm2600 & \citealt{Topping24}\\
\ion{C}{III}]1907,\,1909 & 24.4 & 1.1$^{+0.1}_{-0.2} \times 10^5$ & 24600\pmm2600 & \citealt{Topping24}\\
\ion{Si}{III}]1883,\,1892 & 16.3 & 6.4$^{+5.3}_{-2.6}\times 10^4$ & 24600\pmm2600 & \citealt{Topping24}
\label{tab:ne_results}
\end{tabular}
\end{table*}

The electron density of the highly ionised gas in \RXC can be derived using two independent line diagnostics: \ArIV{4713}/\ArIV{4742} and \OIII{5008}/\OIIIalma. While \OIII{5008}/\OIIIalma depends on both the electron temperature and density, the degeneracy can be broken based on the \Te derived with \OIII{4364}/\OIII{5008} (\eg \citealt{Usui25}). Therefore, we could use both ratios to derive \Ne and probe different density regimes within the ISM due to their distinct critical densities.

\subsubsection{Electron density traced by the \ArIVn lines}
\label{subsub:ArIV_den}

The \ArIV{4713}/\ArIV{4742} doublet probes the high-density and more highly ionised regions \citep{Kewley19}. The \ArIV{4742} line has a critical density about ten times larger than \ArIV{4713} (\ie log($n_\mathrm{c}$[\cubiccm{}])\eqq5.1 and 4.2, respectively), while they both require photons with energies $>$40.7\,eV and therefore trace dense and highly ionised \ion{H}{II} or AGN regions. In addition, by construction, all \Ne-sensitive lines are very close and thus not affected by extinction.

Considering the \HeI{4714} flux values derived assuming the `primordial' and `overabundant' scenarios (see Sect.~\ref{subsec:deblending}), we derived a \ArIV{4713} flux of (6.1\pmm1.4) and (5.8\pmm1.4)\EXP{-20}\ergscm and therefore log(\ArIV{4713}/\ArIV{4742})\eqq-0.06\pmm0.03 and -0.09\pmm0.04, respectively. We then used \texttt{Pyneb} to compute the electron density compatible with these line ratios and the electron temperatures for the `primordial' and `overabundant' He\,I scenarios (see Sect.~\ref{subsec:deblending}). We used the Monte Carlo technique to sample the parameter space with 5000 simulations, obtaining a mean and standard deviation value of log(\Ne{}[\cubiccm{}])\eqq3.8\pmm0.4 and 3.9\pmm0.4 for the `primordial' and `overabundant' scenarios, respectively. We obtained similar results from both scenarios, but we found the \OIIIn-derived \Ne value is more compatible with the `primordial' scenario (see Sect.~\ref{subsub:OIII_den}), and therefore we adopted log(\Ne{}[\cubiccm{}])\eqq3.8\pmm0.4 as the fiducial value for \ArIVn. Figure~\ref{fig:OIII_ne} shows how the \ArIV{4713}/\ArIV{4742} ratio is almost constant at \Te{}\,$>$\,15\,000\,K, minimising the impact of the adopted \Te.

\subsubsection{Electron density traced by the optical and far-IR \OIIIn lines}
\label{subsub:OIII_den}

\begin{figure}
\centering
   \includegraphics[width=\linewidth]{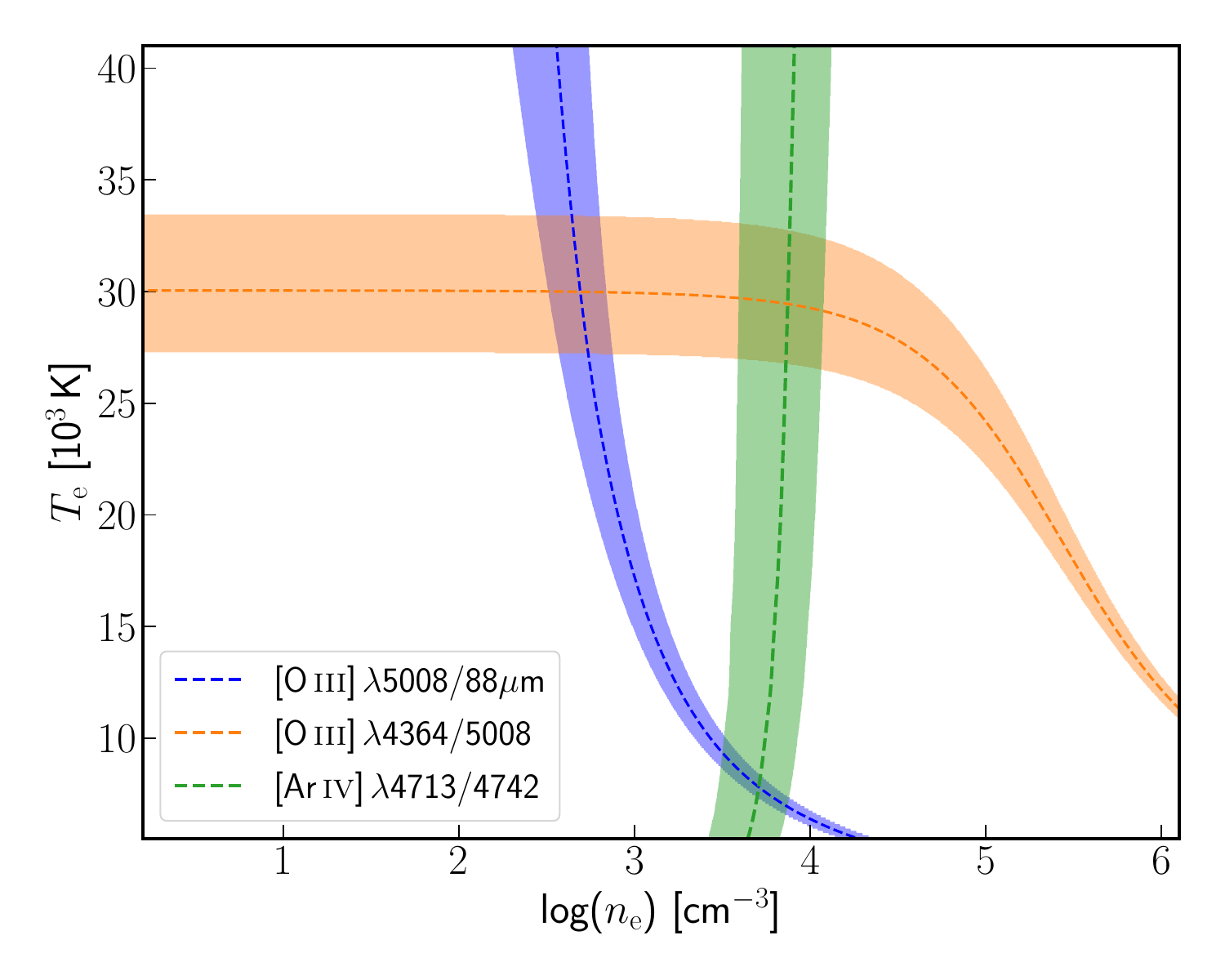}
      \caption{Electron density and temperature diagnosis. The green line and shadowed area show the ratio \ArIV{4713}/\ArIV{4742} and its uncertainty derived after deblending the \HeI{4714} contribution (see Sect~\ref{subsec:deblending}). Similarly, orange and blue lines and shadowed areas represent the \OIII{4364}/\OIII{5008} and \OIII{5008}/\OIII{88} ratios which trace the \Te and \Ne, respectively.}
         \label{fig:OIII_ne}
\end{figure}

The detection of the \OIIIalma line with ALMA allowed us to determine the electron density of the ionised O$^{++}$ emitting gas using the \OIII{5008}/\OIIIalma line ratio, a widely used diagnostic for ionised gas due to their moderately high ionisation potential ($>$35.1\,eV; \eg \citealt{Stiavelli23, Harshan24, Fujimoto24, Harikane25, Usui25}).

The critical density of \OIII{5008} (\ie{} log($n_\mathrm{c}${}[\cubiccm{}])\,\simi5.8) is a about three orders of magnitude larger than for \OIIIalma (\ie{}log($n_\mathrm{c}${}[\cubiccm{}])\,\simi2.7), which makes this ratio very sensitive to electron density over a wide dynamic range. However, their large distance in wavelength causes this ratio to be very dependent of attenuation, as the \OIII{5008} line can potentially be affected by dust extinction.

For this analysis, we only considered the \total flux for \OIII{5008}, as the ALMA beam size does not allow us to spatially resolve the different clumps in \RXC. In addition, the ALMA \OIIIalma FWHM (\ie 174\pmm43\kms; see Sect.~\ref{subsec:ALMA}) suggests the detection of the same broad component observed in the optical \OIIIn line (see Sect.~\ref{subsec:Kin}). Therefore, in this section, we use both the narrow and broad component fluxes derived for \OIII{5008}. After correcting the broad emission in the optical \OIIIn from dust attenuation (\Av{}\eqq1.5\pmm0.5, see Sect.~\ref{subsec:extinction}), we derived log(\OIII{5008}/\OIIIalma)\eqq1.5\pmm0.1.

As in the previous section, we made use of the theoretical \texttt{Pyneb} ratios to derive the electron density
compatible with our \OIII{5008}/\OIIIalma value and the [O$^{++}$] electron temperature computed in Sect.~\ref{subsec:temp_OIII}. We performed 5000 Monte Carlo simulations to sample the parameter space covered by the measured \OIII{5008}/\OIIIalma and \OIII{4364}/\OIII{5008} ratios, obtaining a mean value of log(\Ne{}[\cubiccm{}])\eqq2.7\pmm0.2. Figure~\ref{fig:OIII_ne} shows the electron temperature and density diagnosis based on these line ratios.

The electron density we obtained is in agreement with the value found by \citet{Yanagisawa24_He} from the He\,I lines assuming a primordial helium abundance (\ie \Ne{}\eqq1128\oddpm{1124}{407}\cubiccm), thus favouring this scenario. For the lower electron densities expected in the He\,I-overabundance scenario (\Ne{}$<$\,50\cubiccm), we would need a much lower \OIII{5008}/\OIIIalma ratio than the one measured, as the \OIIIalma line emission would be boosted in regions with densities lower than its critical density ($n_\mathrm{c}$\eqq510\cubiccm).

The density derived from the \OIII{5008}/\OIIIalma ratio \Ne{}\simi500\cubiccm{}) should be interpreted as an emission-weighted measure of the O$^{++}$ gas phase. This value indicates that while denser compact clumps may be present, the bulk of the O$^{++}$ emitting gas is characterised by moderate densities. Interestingly, this density is significantly lower than the \Ne{}\simi10$^4$\cubiccm{} expected at \z{}\eqq6 from high-ionisation tracers such as \ion{C}{III}] (e.g. \citealt{Topping25,Martinez25}) and aligns more closely with those typically inferred from low-ionisation diagnostics, such as [\ion{O}{II}] and [\ion{S}{II}] (e.g. \citealt{Abdurrouf24,Topping25}). Consequently, this suggests that despite its high-ionisation state, the O$^{++}$ emission in \RXC is dominated by a diffuse and extended gas component rather than the high-pressure environments typically associated with high-ionisation species.

\section{Discussion}
\label{sec:Discussion}

\subsection{The complex multi-phase ISM structure in RXCJ2248-ID: A common feature amongst high-\z extreme N-emitters?}
\label{subsec:ISM_struct}

\begin{figure}
\centering
   \includegraphics[width=\linewidth]{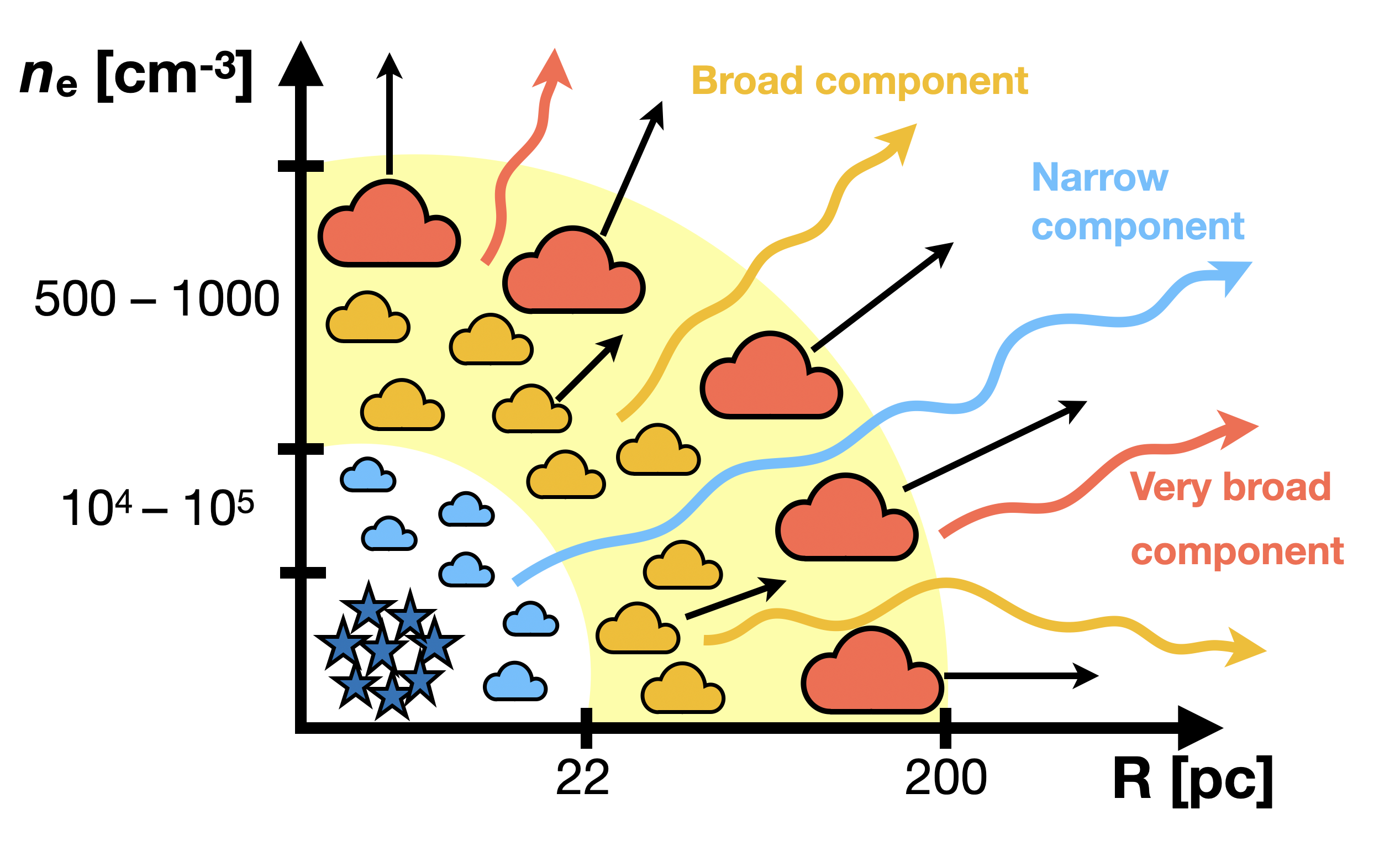}
    \caption{Illustration of the model proposed assuming a generally spherical but anisotropic distribution. Blue, gold and red clouds represent three different ISM layers that we linked to the kinematic components found in Sect.~\ref{subsec:Kin}. In addition, we include a more diffuse component (in yellow) corresponding to \OIIIalma emission (\Ne{}\simi500\cubiccm). In this model, the narrow component is emitted from the inner and more dense regions (in blue), while the dusty outflowing gas lie at larger distances (gold and red).}
    \label{fig:model_rad}
\end{figure}

RXCJ2248-ID belongs to the remarkable class of high-\z galaxies identified as extremely compact and extreme N-emitters with an enhanced nitrogen abundance \citep{Naidu26}. Its physical size and star formation and stellar surface density ($r_\mathrm{eff}$\,$<$\,22\,pc, $\Sigma_\mathrm{SFR}$\,$>$\,10400\Msyr{}\,kpc$^{-2}$ and $\Sigma_\mathrm{*}$\,$>$\,3.6\EXP{4}\,\Msun{}\,pc$^{-2}$; \citealt{Topping24}) are in the range with what has been measured in other extremely high-\z (\z{}$>$8 and up to 14.4) N-emitters, such as Gz9p4 \citep{Schaerer24}, CEERS-1019 \citep{Marques-Chaves24}, GNz11 (\citealt{Bunker23,Maiolino_24,Alvarez-Marques25,Crespo-Gomez26}), GHz2 \citep{Calabro24,Castellano24,Zavala24,Zavala25}, and MoM-z14 \citep{Naidu26}.

The combination of our new NIRSpec and ALMA data together with existing results \citep{Topping24, Yanagisawa24_He} shows that the ionised ISM in RXCJ2248-ID has a complex non-uniform structure characterised by a range of physical conditions and kinematics in a very small and compact region. While the UV lines (\eg \ion{Si}{III}], \ion{N}{IV}]) trace a highly dense medium (6.4\EXP{4}\,$\leq$\,\Ne{}(\cubiccm{})\,$\leq$\,3.1\EXP{5}), the optical and far-IR lines indicate the presence of a less dense medium with densities of \Ne{}\,\simi6.7\EXP{3}\cubiccm and down to \simi500\cubiccm, as traced by the \ArIVn and \OIIIn lines, respectively (see Table~\ref{tab:ne_results}).  These differences are naturally explained by the different critical densities of the transitions, which determine the density regimes in which each line is most efficiently emitted. The UV doublets, with higher critical densities, preferentially trace the densest clumps within the ionised medium, while \ArIVn, with a lower critical density, is more sensitive to lower-density gas within the same overall high-ionisation region \citep{Kewley19}. Moreover, while the ISM shows an extreme range in density (factor of 600 between the dense clumps and diffuse medium), the temperature is very high (\simi25000\range30000\K), even for low-metallicity galaxies \citep{Sanders20,Isobe22,Curti23,Morishita24,Rickards-Vaught25}, independent of the ionisation status and electron density of the gas.

A physical picture emerges from these results in which the structure of the ISM is clumpy and
non-uniform. The ISM in \RXC would
consist of gas clouds of different densities within a more diffuse and tenuous medium, as already observed in other high-\z galaxies \citep{Harikane25,Usui25} and
predicted by cosmological simulations \citep{Choustikov26,Nakazato26}. Under this scenario, depicted in Fig.~\ref{fig:model_rad}, high-density clouds (in blue; \Ne{}\simi10$^4$\range 10$^5$\cubiccm{}) would dominate the emission of the strong UV collisional  lines close to the compact UV-bright region (\simi22\,pc). The lower density gas (yellow region; \Ne{}\simi500\cubiccm{}) would emit a major fraction of the \OIIIalma flux at larger distances, \simi200\,pc to \simi750\,pc, as given by the NIRSpec and ALMA angular resolutions, respectively. The optical lines (\eg \OIII{5008}, \Ha) would be generated in high-density clouds and diffuse gas, covering the entire range of densities derived in Sect.~\ref{subsec:density}.

The volume occupied by the high- and low-density gas can be estimated from the \OIII{5008} and \OIIIalma luminosities, assuming densities of 10$^5$ and 500\cubiccm{}, respectively. Following \citet{Usui25}, the very dense gas occupies a very small region ($r_\mathrm{hd}$\simi2.5\,pc), while the more diffuse component is distributed over a larger area ($r_\mathrm{ld}$\simi120\,pc). These sizes imply volume filling factors of 0.2$\%$ and 10$\%$ for the high- and low-density gas, assuming emission regions with a size of $r$\eqq20 (compact UV-bright) and 200\,pc (NIRSpec unresolved size), respectively. The electron temperature of all of these clouds appears to  be high (\Te{}$>$25000\K), likely as a consequence of the low-metallicity of the galaxy (\ie less cooling due to the metal lines) and the strong ionising radiation field (log($U$)\,\simi$-1$) resulting from the extreme star formation surface density above 10400\,$M_{\odot}$\,yr$^{-1}$\,kpc$^{-2}$ \citep{Topping24}. In this scenario, the broad an very broad velocity components of the ionised gas  (see Sect.~\ref{subsec:Kin}) would correspond to the outflows of some of these dusty clouds. Since the filling factor of the ionised gas is low, the UV radiation from the central starburst would not be much affected by the dusty outflowing clouds and would appear with a low extinction. Even though the ISM structure shown in Figure~\ref{fig:model_rad} explains the multi-phase ionised gas and the three velocity components, it still has difficulties explaining the slightly higher $A_\mathrm{V}$ of the very broad component relative to the broad component. To make the very broad component appear redder, the optical path would need to be longer. Therefore, more complex structures, possibly induced by turbulence during the outflow, might be required.

A multi-layered ISM with a compact and very high-ionisation central zone has also been found in metal-poor extreme emission-line galaxies (\citealt{Berg2021}). In fact, the scenario proposed in this work agrees with the ISM stratification found in the COS Legacy Archive Spectroscopic SurveY (CLASSY), which revealed that UV-based electron densities are about two orders of magnitude larger than those derived from optical lines in high-\z analogues and cover a range similar to that measured in RXCJ2248-ID \citep{Mingozzi22}.
A similar result has been observed in the far-IR regime, where \citet{Lebouteiller12} found that \OIIIalma emission is photoionised by far-UV photons that penetrate low-density regions due to a clumpy ISM structure. In fact, low-metallicity has been found to be linked to an increment in the ISM porosity, favouring a clumpy structure in the gas and dust clouds where the hard ionisation field can escape at larger distances than in more homogenous ISMs \citep{Izotov16,Cormier19,Plat19}.

\subsection{Dusty outflows in RXCJ2248-ID: Attenuation-free and feedback-free star-forming scenarios} 
\label{subsec:outflows_discussion}

In addition to the density structure and clumpiness of the ISM, the new NIRSpec data prove that the ISM also has a very complex velocity structure. Aside from the confirmation of the very broad (FWHM\,\simi1200\kms) component previously detected \citep{Topping24}, we identified a very narrow (systemic) component (FWHM\,\simi65\kms) and a new nuclear broad (FWHM\,\simi230\kms) emission line centred on the systemic velocity. The very blue UV continuum slope of \RXC ($\beta$\eqq$-$2.72, \citealt{Topping24}) and the Balmer decrement found for the systemic component (\Ha/\Hb{}\,\simi2.70; see Sect.~\ref{subsec:extinction}) are compatible with no attenuation in both the UV-bright stellar population and associated ionised gas phase. However, we find that the broad and very broad components of the ionised gas have extinctions (\Av{}\eqq1.5 and 2.5\,mag, respectively). Interestingly, the very broad component velocity width varies between \Ha{} (\simi1000\kms) and \OIIIn (\simi1500\kms). This kinematic stratification is common in energetic outflows, where the lower-density, highly ionised gas traced by $\text{[O III]}$ is more efficiently accelerated in the presence of an AGN \citep{Harrison16,Kang17,Ubler23}. Additionally, the higher dust extinction affecting the broad nuclear component may partially obscure the highest-velocity gas of the densest phase, leading to a narrower observed profile for \Ha{} compared to \OIIIn. Therefore, we interpret these broad components as stratified dusty outflows originating in the most inner regions of RXCJ2248-ID and blowing the dust from these regions with maximal velocities of \simi115\range500\kms for \Ha ($v_\mathrm{max}$\eqq$|\Delta V+\mathrm{FWHM}/2|$\ , \citealt{Arribas14}).

We derived the total outflowing ionised gas mass using Equations 5 from \citet{Carniani24} based on the \Ha dust-corrected luminosities, gas metallicity (0.05\Zsun; \citealt{Topping24}), and electron densities. Assuming the \Ne derived for the \ArIVn emitting gas (\simi6700\cubiccm) and the extinction-corrected \Ha fluxes for the broad and very broad line components (7.5 and 5.9\EXP{42}\ergs, respectively), we estimate a total gas mass involved in the outflows of 1.6\EXP{6}\Msun. If the outflowing gas is more diffuse and has an electron density close to the \Ne value derived for the \OIIIn (\ie 500\cubiccm{}), the outflow mass would be larger by a factor of 13 (\ie 2.1\EXP{7}\Msun).

A certain amount of this outflowing mass may be able to escape into the intergalactic medium, with subsequent metal enrichment, while another fraction would return to the galaxy, forming stars at a later time. An estimate of these masses can be obtained by comparing the outflowing and escaping velocities. 
For \RXC, the escape velocity is derived from its dynamical mass, assuming a range of sizes for the halo relative to that of the compact starburst ($r_\mathrm{eff}$\eqq22\,pc). Following \citet{Arribas14}, the escape velocity is in the range of 62\range150\kms for halos with sizes 1 to 100 times $r_\mathrm{eff}$ and the dynamical mass derived previously (see Sect.~\ref{subsec:dynamicalmass}). Therefore most of the ionised outflowing mass detected in \RXC would have velocities high enough ($v_\mathrm{max}$\,\simi115\range500\kms) for the gas and dust to escape the galaxy gravitational potential and contribute to the chemical enrichment of not only the external ISM but also the circum- and intergalactic medium. This would be in agreement with previous findings for other high-\z galaxies \citep{Carniani24}. Note, however, that if the SED-based stellar mass is considered (see discussion in Sect.~\ref{subsec:dynamicalmass}), the escape velocity would be larger by a factor of two to three (i.e. up to 450\kms), and therefore only a small fraction of the outflowing gas would be able to escape the galaxy. Under this scenario most of the outflowing gas returns to the central regions, forming stars at a later time (e.g. fountain-like scenario). A more accurate derivation of the total mass of the system is required before reaching any firm conclusion about the evolution of the outflowing material in \RXC. 

The mass loading factor ($\eta$\eqq$\dot{M}$/SFR) can be estimated from the derived masses of the outflowing ionised gas, the maximal velocities of these outflows, and the SED-derived SFR (\ie63\Msyr; \citealt{Topping24}). For the range of masses derived asumming different electron densities, we obtained $\eta$\,\simi0.3\range4, which is compatible with other high-SFR galaxies at \z{}\,>\,2 \citep{Davies24,Rodriguez-delPino24,Ren25,Xu25} and lies on the lower end of the range derived in other log(\Mstar/\Msun)\,<\,8 high-\z JADES galaxies ($\eta$\,\simi2\range12; \citealt{Carniani24}). This suggest that the outflows in \RXC traced by the ionised gas phase, although significant, might not be strong enough to quench the star formation in this extreme starburst.

One of the models to explain the existence of extremely UV-bright and compact high-\z starbursts, such as in \RXC, is the so-called attenuation-free model (AFM; \citealt{Fiore23,Ferrara23, Nakazato25}).  
In this model, the dust is produced by early supernovae, but it is rapidly removed from star-forming regions by radiation pressure and stellar feedback-driven dusty outflows. The detection of broad and very broad line components with visual extinctions of 1.5 and 2.5 magnitudes, together with the evidence of no extinction in the narrow component, supports this scenario.

Intense star formation in the nuclear region increases the UV photon output from young stars. The UV radiation is absorbed by dust grains, which gain momentum from the radiation pressure and transfer it to the surrounding gas. As a result, a dusty outflow, dynamically coupled with the dust and gas, is launched. This outflow removes gas and dust from the central regions, allowing the UV emission to escape.
Specifically, \citet{Ferrara23} analytically derived the threshold specific-SFR (sSFR) for the onset of dusty outflows as sSFR\,$>$\,25\,Gyr$^{-1}$. 
Assuming the stellar mass (log(\Mstar/{}\Msun)\eqq8.05) and SFR (log(SFR/\Msyr)\eqq1.8) presented in \citet{Topping24}, \RXC exceeds this condition with an extreme sSFR value of \simi550\,Gyr$^{-1}$. Recent hydrodynamical simulations of stellar cluster formation have demonstrated that compact starbursts ($\Sigma_\mathrm{*}$\,$>$10$^{3}$\Msun\,pc$^{-2}$) can expel the surrounding gas and dust through radiation-driven outflows up to tens of parsecs \citep{Menon24}. These allow the intense UV-emission to escape from the compact starburst, while more external clouds regions can still be obscured, yielding a stratification and/or clumpiness of the ISM, as presented in Fig.~\ref{fig:model_rad}. In fact, recent works have suggested that radiation-driven outflows are causally linked to extremely compact SFGs \citep{Dessauges-Zavadsky25,Marques-Chaves25,Vanzella26}. The AFM scenario might simultaneously explain the observed dusty outflows with different degrees of attenuation and velocities.

In other models, such as the feedback-free starburst (FFB) scenario proposed by \citet{Dekel23}, a dense gas cloud collapses within a few megayears before stellar feedback (winds and SNe) becomes effective, leading to intense star formation with high star formation efficiency. The electron density measured from the UV and \ArIVn emission lines is well above the critical density (3000\cubiccm{}) needed to ensure a short ($\leq$\,1\,Myr) free-fall \citep{Dekel23}. The lower end of the mass loading factor ($\eta$\eqq0.3) and outflowing velocities (\simi115\kms) ranges derived for \RXC also supports this scenario, where the outflowing gas cannot escape the central regions and will later produce a new generation of stars during a 100 Myr long FFB phase or an inflow of gas feeding a black hole seed in a post-FFB period \citep{Li24}. 
Winds driven by SNe from post-FFB star clusters of previous generations are expected to be observed at FWHM\,$>$\,1000\kms \citep{Li24}. Although the outflows traced by the very broad \OIIIn and \Ha line components show FWHM values consistent with typical SNe-driven velocities suggested by analytical calculation \citep{Lagos13} and hydrodynamical simulations \citep{Li17}, the estimated age of \RXC (\simi2\Myr, \citealt{Topping24}) would be too young for SNe to have occurred under normal assumptions. However, the recent detection of Wolf--Rayet features by \citet{Berg26} suggests a somewhat older starburst age of \simi3\range6\Myr, at which the onset of supernovae from the most massive stars becomes plausible. Thus, SNe-driven feedback cannot be excluded as the origin of the observed outflows. Alternatively, UV- and/or radiation-driven feedback may contribute to driving the outflows \citep{Ferrara23,Ferrara25}. More extreme stellar populations, including very  and super massive stars (>\,100\Msun), could provide an additional route to very early feedback, and they have recently been suggested to explain the high N/O ratio detected in some high-\z galaxies \citep{Charbonnel23,Ji24,Vink25,Zhang26}.
The existing data covering up to 0.7\micron, however, do not provide evidence of an additional mature stellar population that could have produced a previous generation of supernovae.

Regardless of the origin of the powering source, this study shows the presence of a complex ISM with different densities and the existence dusty outflows with different velocities and levels of attenuation in \RXC. As we understand from this work, these outflows blow the gas and dust far outside the very compact and dense star-forming regions, revealing their extreme UV-emission, high electron densities, and peculiar N/O ratios. This interpretation agrees with the results from the dusty outflows found in another compact N-enhanced galaxy at \z{}\,\simi6 \citep{Marques-Chaves25}, suggesting the existence of a population of galaxies with similar physical properties. Further analysis with deep spectroscopy and imaging that also extends into the optical-red and near-infrared are required to understand the nature of \RXC and ultimately the nature of N-emitter galaxies.

\section{Summary and conclusions}

\label{sec:5.Summary}

We have presented new NIRSpec/IFU data for the magnified ($\mu$\eqq5.3) N-enhanced galaxy \RXCthree (\z{}\eqq6.105) at high spectral resolution (R\,\simi2700) that allowed us to look for potential spatial and kinematic differences in its physical properties. Concretely, we used the detected optical emission lines to study the kinematic and dust distribution of the substructure of its ionised ISM as well as its electron temperature and density. In addition, we combined NIRSpec/IFU with ALMA to derive the optical-to-far-IR \OIII{5008}/\OIIIalma line ratio that is used to estimate the electron density of the doubly ionised oxygen regions. The main results drawn from this work are summarised as follows:

\begin{itemize}
    \item We identified three distinct kinematic components in the brightest UV clump using the \OIII{5008} and \Ha{} lines with intrinsic an FWHM of \,\simi60\range70, \simi210\range250 and \simi1000\range1500\kms. The presence of three kinematically distinct components indicates a complex kinematic substructure in a very compact region (\simi22\,pc).
    
    \item The ionised ISM has a complex and non-uniform dust distribution, as traced by the (\Ha/\Hb) ratio derived for the three different kinematic line components.
    While the narrow component is compatible with negligible dust extinction, the broad and very broad line components are dusty, with visual extinctions \Av{} of 1.5 and 2.5 magnitudes, respectively. This is compatible with a scenario where the broad components trace  dusty outflows very close to the central regions of the extinction-free UV-bright burst.

    \item The high ionisation [OIII] emitting gas has a very high electron temperature of \Te{}\,\simi30000\K, which is slightly larger than previous electron temperatures derived using the UV lines. This temperature is higher ($\times$1.5\range2) than temperatures measured in low-\z low-metallicity galaxies. We suggest that this high \Te is linked to the presence of a  harder ionising source in \RXC.

    \item We identified a clumpy ISM with a wide range of electron densities. The electron density of the ionised gas traced by the ratio of the \ArIV{4713}/\ArIV{4742} doublet (log(\Ne{}[cm$^{-3}$])\eqq3.8\pmm0.4) is more than an order of magnitude smaller than the density previously found using UV doublets (\ie \ion{Si}{III}], \ion{C}{III}], \ion{N}{IV}]). We also obtained a lower value based on the optical-to-far-IR \OIII{5008}/\OIIIalma ratio (\ie log(\Ne{}[cm$^{-3}$])\eqq2.7\pmm0.2). The volume occupied by the high- and low-density clouds is small and corresponds to filling factors of about 0.2\% and 10\%, respectively. 

    \item We propose a scenario where the UV continuum and high-ionisation line emission is produced in a very compact region (\simi22\,pc), while the optical and far-IR \OIIIn is emitted by less dense clouds (\simi500\range1000\cubiccm{}) in more extended regions (200\range750\,pc). In this scenario, dusty gas clouds are outflowing with maximal velocities $\Delta V$ of \eqq115\range500\kms, clearing the inner regions from dust and allowing the strong central UV emission to be observed free of extinction.

    \item The velocities of the dusty outflows are such that a fraction of the total outflowing mass (0.16\range2.1\EXP{7}\Msun), in particular the high-velocity component, will escape into the intergalactic medium, supporting the AFM. The rest of the outflowing mass will fall back to the central region and would be available for additional star formation. On the other hand, the high densities measured in the ionised ISM are above the critical densities needed for a short free-fall time. This, together with the lower end of the mass loading factor (\simi0.3) and gas outflowing velocities (\simi115\kms), qualitatively supports the FFB model with gas available for short bursts over a period of 100 Myr and for feeding of post-FFB black hole seeds. Additional JWST data covering the rest near-infrared range would be required to identify the presence of mature stellar populations (100\Myr) or hot dust emission associated with small quantities of dust in the central UV-bright region, as expected in the FFB and AFM scenarios, respectively.

\end{itemize}

In summary, the analysis carried out for \RXC using NIRSpec/IFU high-resolution and ALMA \OIIIalma data allowed us to disclose the extremely complex structure of the ionised ISM in the compact UV-bright galaxy \RXC. In particular, with the new data we identified the presence of dusty outflows with different velocities and internal extinction. In addition, the new data reveal a range of physical conditions in the compact UV-bright regions characterised by a low filling factor, an extremely high electron temperature ($\times$1.5\range2 larger than low-\z low-metallicity galaxies) and a wide range of electron densities, spanning from low (500 cm$^{-3}$) to intermediate (6700 cm$^{-3}$) and previously reported very high values (3\EXP{5}\cubiccm{}). We interpret these results as evidence of an extremely clumpy or stratified ISM, where the different kinematic components and ionised species trace physically distinct regions. In addition, the presence of dusty outflows, which remove the dust from the nuclear regions, could justify the extreme UV-emission observed in \RXC and other compact N-enhanced galaxies at high \z. Some of the characteristics observed in \RXC seem to support the AFM scenario, while others support the FFB model. Further deep JWST and ALMA spectroscopy are needed to further investigate the properties of \RXC and evaluate the theoretical scenarios. In particular deeper ALMA observations with higher angular resolution are required to establish the extent of the low-density \OIIIalma emitting region and whether outflows are detected in the low-density gas. These measurements will provide accurate values for key quantities such as the filling factor, outflowing mass, and mass loading factor.

\begin{acknowledgements}

L.C., J.A.-M. and S.A. acknowledge support by grant PIB2021-127718NB-I00 from the Spanish Ministry of Science and Innovation/State Agency of Research MCIN/AEI/10.13039/501100011033 and by “ERDF A way of making Europe”. 
L.C., J.A.-M., C.B.-P., L.C., S.A. acknowledge support by grant CSIC/BILATERALES2025/BIJSP25022.
T.H. was supported by the Leading Initiative for Excellent Young Researchers, MEXT, Japan (HJH02007) and by JSPS KAKENHI grant Nos. 22H01258, 23K22529, and 25K00020.
C.B.P. acknowledges the support of the Consejer\'ia de Educaci\'on,Ciencia y Universidades de la Comunidad de Madrid through grants No. PEJ-2021-AI/TIC-21517 and PIPF-2023/TEC29505.
J.A.-M. and C. B-P acknowledge support by grant PID2024-158856NA-I00 from the Spanish Ministry of Science and Innovation/State Agency of Research MCIN/AEI/10.13039/501100011033 and by “ERDF A way of making Europe”.
The project that gave rise to these results received the support of a fellowship from the “la Caixa” Foundation (ID 100010434). The fellowship code is LCF/BQ/PR24/12050015.
Y.T acknowledges the support from JSPS KAKENHI Grant Numbers 22H04939, 23K20035, and 24H00004.
Y.N. acknowledges funding from JSPS KAKENHI Grant Number 23KJ0728 and a JSR fellowship.
D.C. is supported by research grant PID2021-122603NB-C21 funded by the Ministerio de Ciencia, Innovaci\'{o}n y Universidades (MI-CIU/FEDER) and the research grant CNS2024-154550 funded by MI-CIU/AEI/10.13039/501100011033.
M.H. is supported by Japan Society for the Promotion of Science (JSPS) KAKENHI Grant No. 22H04939.
K.M. acknowledges financial support from the Japan Society for the Promotion of Science (JSPS) through KAKENHI grant No. 20K14516. 
K.M. and A.K.I are supported by JSPS KAKENHI grant No. 23H00131.
N.Y. acknowledges financial support from JSPS Specially Promoted Research 24H00004

This work is based on observations made with the NASA/ESA/CSA James Webb Space Telescope. The data were obtained from the Mikulski Archive for Space Telescopes at the Space Telescope Science Institute, which is operated by the Association of Universities for Research in Astronomy, Inc., under NASA contract NAS 5-03127 for JWST. These observations are associated with program $\#$1840.

This paper makes use of the following ALMA data: ADS/JAO.ALMA$\#$2013.1.01052.S. ALMA is a partnership of ESO (representing its member states), NSF (USA) and NINS (Japan), together with NRC (Canada), NSTC and ASIAA (Taiwan), and KASI (Republic of Korea), in cooperation with the Republic of Chile. The Joint ALMA Observatory is operated by ESO, AUI/NRAO and NAOJ.

\newline
\newline

\textbf{Data availability:}
The data used in this work may be obtained from
\url{https://dx.doi.org/ 10.17909/rq91-kw87}.

\end{acknowledgements}

\bibliographystyle{aa} % style aa.bst
\bibliography{bibliography.bib} % your references Yourfile.bib

\begin{appendix}

\section{ALMA observations}
\label{ap1:ALMA}

\begin{table}[h!]
\caption{ALMA Band 8 observations.}\label{tab:ALMA-shogen}
\centering
\begin{tabular}{lcccccc} % 7 cols.
\hline
\hline
Parameter && \multicolumn{5}{c}{\bf Band 8} \\
\cline{1-1}
\cline{3-7} 
Frequency (GHz)\tablefootmark{$\dag$} & & \multicolumn{5}{c}{465.23, 467.04, 477.25, 479.12}\\
Date & & 2015-05-02 & 2015-05-16 & 2015-06-06 & 2016-05-18 & 2016-08-20\\
No.\ of antennas & & 35 & 41 & 38 & 42 & 45 \\
Baseline length (m) & & 14--356 & 15--558 & 21--820 & 14--626 & 15--704 \\
Correlator mode & & \multicolumn{5}{c}{Frequency division mode}\\
On-source time (min) & & \multicolumn{5}{c}{ 292.5 }\\
PWV (mm)& & \multicolumn{5}{c}{0.39--0.50} \\
Flux calibrator     & & Ceres & Titan & Ceres & Pallas & J0006$-$0623\\
Gain calibrator     & & \multicolumn{5}{c}{J2230$-$4416, J2235$-$4835}\\
Bandpass calibrator & & \multicolumn{5}{c}{J1924$-$2914, J2235$-$4835, J0006$-$0623}\\
\hline
\end{tabular}
\tablefoot{
\tablefoottext{$\dag$}{The centre frequencies of correlator spectral windows.}
}
\end{table}

\begin{figure}[h!]
    \includegraphics[width=0.48\textwidth]{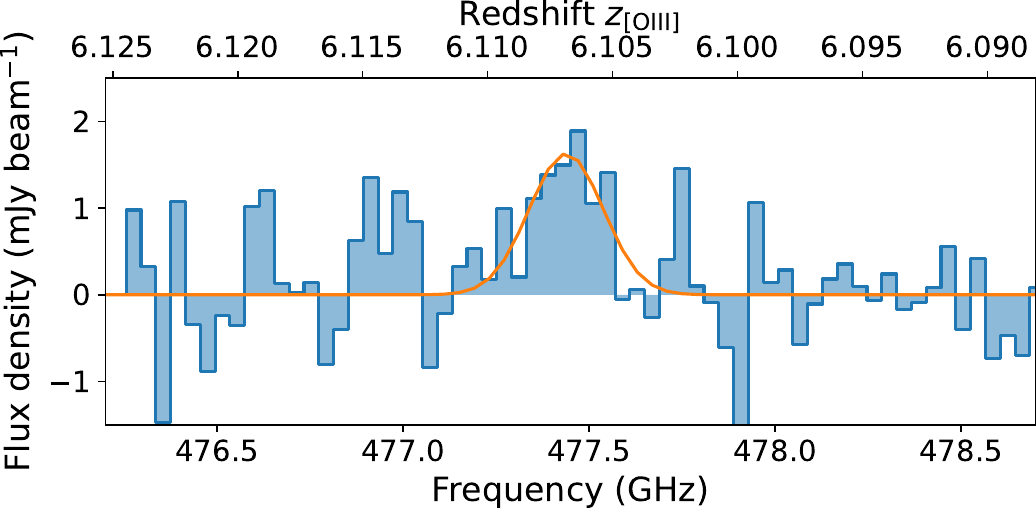}
    \caption{ALMA \OIIIalma spectrum measured at the peak position of RXC~J2248-ID3. The upper horizontal axis shows the redshift for the \OIIIalma line. The orange curve is the best-fitting Gaussian.}
    \label{fig:alma_spec}
\end{figure}

\section{Complementary line fitting figures}
\label{ap2:line-fit}

Here we present complementary figures with the line fitting analysis for the \OIII{4960}+\OIII{5008 and \Hb lines, as well as the \ArIVn doublet.}
\begin{figure*}[h!]
    \centering
    \includegraphics[width=0.9\textwidth]{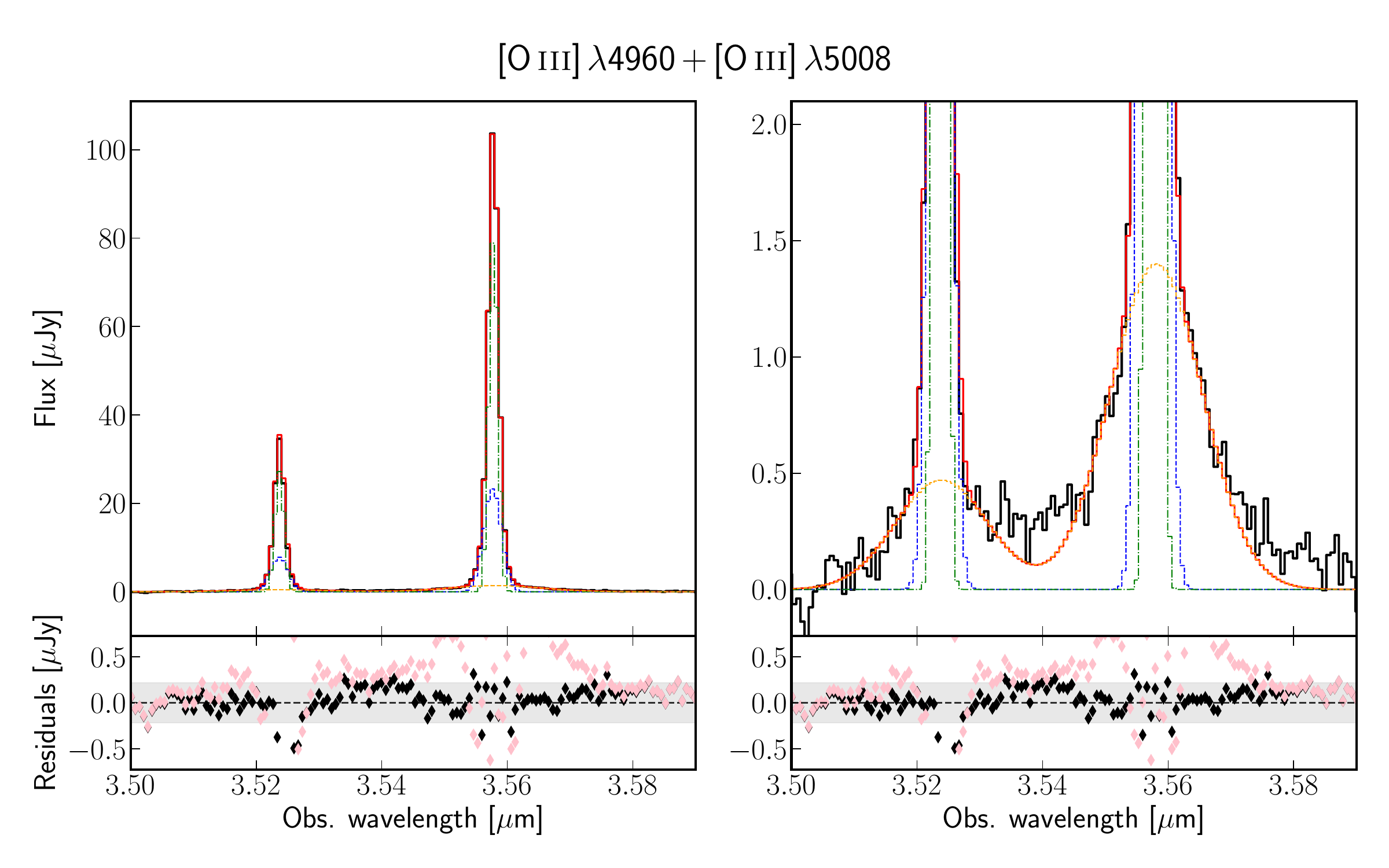}
    \caption{Simultaneous fit of \OIII{4960} and \OIII{5008} as described in Sect.~\ref{subsec:int_spec}. Solid black and red lines represent the observed flux and the best-fit model considering three independent Gaussians for each line. Dot-dashed green, and dashed blue and orange dashed lines display the narrow, broad and very broad components. Bottom panels present the residuals from the best-fit model considering three and two Gaussian profiles, respectively. The right panel shows a zoom-in of the left panel to better visualise the contribution of the fainter components.}
    \label{fig_ap:o3_fir}
\end{figure*}

\begin{figure}[h!]
    \includegraphics[width=0.45\textwidth]{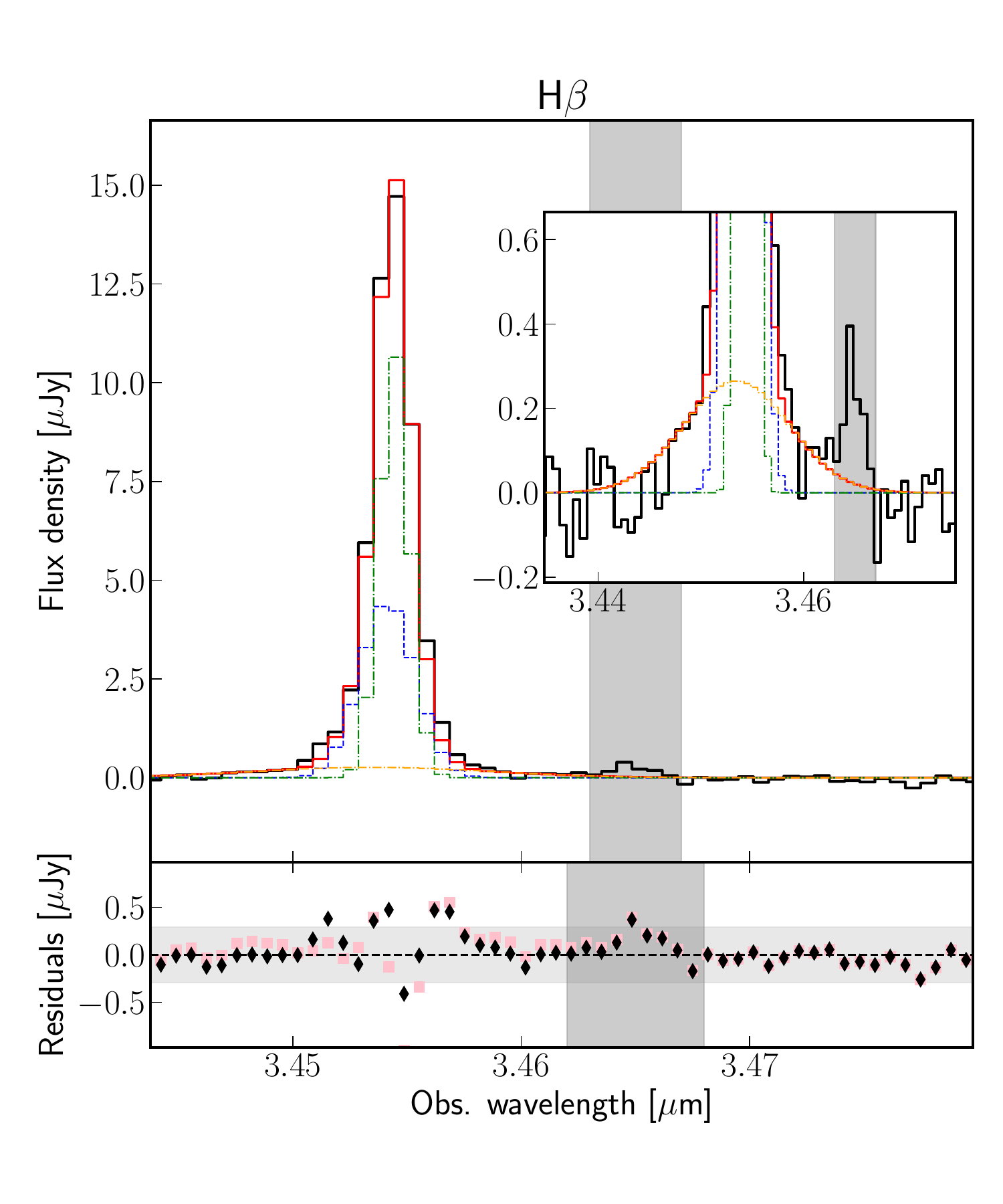}
    \caption{Same as Figure~\ref{fig:Broad_fitfig}, but shown for \Hb. The three-Gaussian fitting was performed simultaneously with \Ha{}\,+\,[\ion{N}{II}] (see Sect.~\ref{subsec:int_spec}). Note that, although we obtain similar residuals using two and three Gaussian profiles for \Hb, the simultaneous \Ha{}\,+\,\Hb fit considering three components show a lower AIC.}
    \label{fig_ap:Hb}
\end{figure}

\begin{figure}[h!]
    \includegraphics[width=0.45\textwidth]{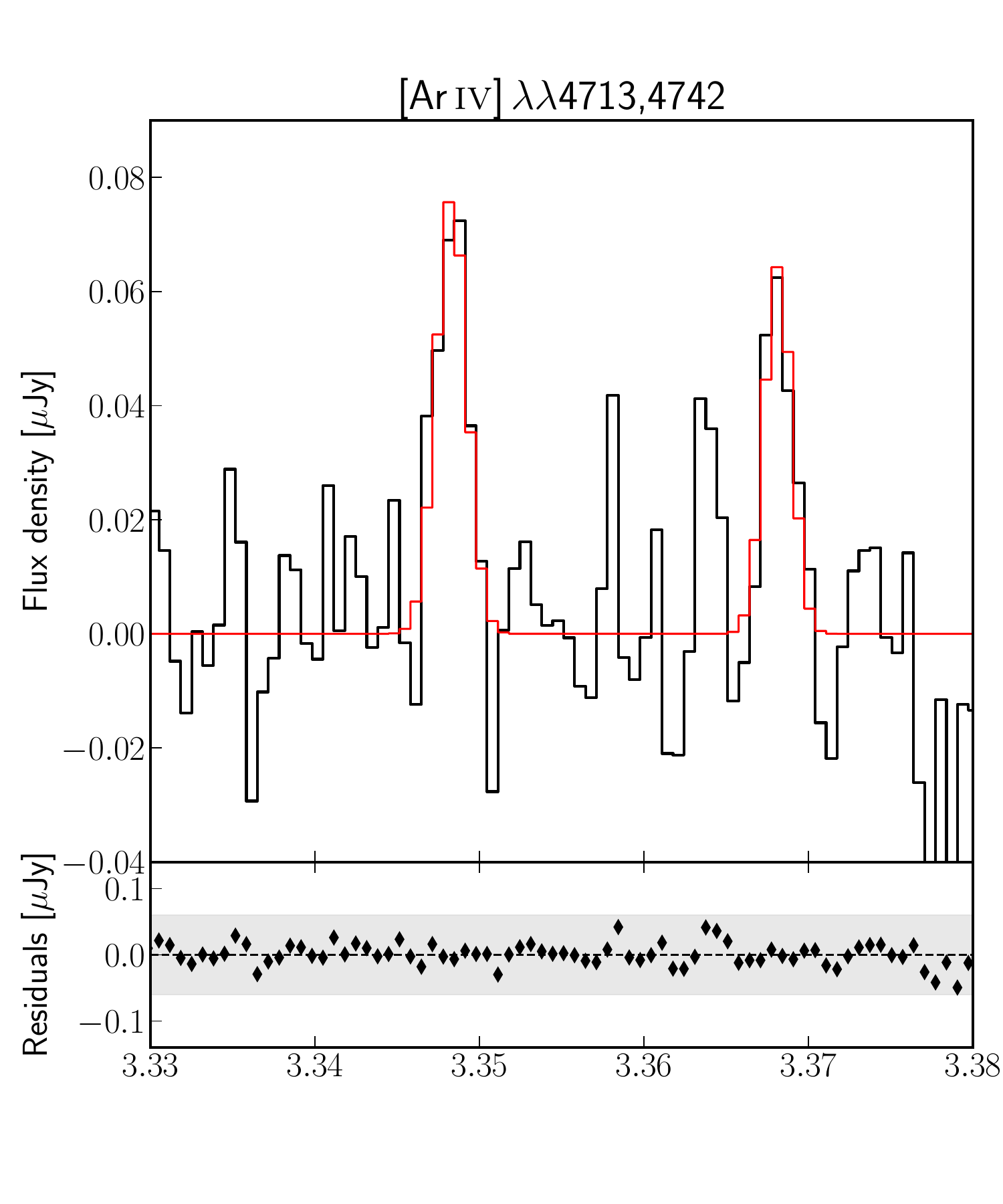}
    \caption{Single-Gaussian fitting for the \ArIVn doublet from the nuclear spectra. Note that the \ArIV{4713} line is blended with \HeI{4714} (see Sect.~\ref{subsec:deblending}).}
    \label{fig_ap:Ar}
\end{figure}

\end{appendix}

\end{document}